\documentclass{aa}

\usepackage{graphicx}
\usepackage{txfonts}
\usepackage{caption}
\usepackage{subcaption}
\usepackage{xcolor}
\usepackage{url}
\usepackage{hyperref}
\usepackage{placeins}
\usepackage{comment}

\begin{document}

   \title{Twenty years of blazar monitoring with the INAF radio telescopes}
   \author{N. Marchili\inst{1}
          \and
          S. Righini\inst{2}
          \and
          M. Giroletti\inst{1}
          \and
          C.M. Raiteri\inst{3}
          \and
          R. P. Giri\inst{1,4}
          \and
          M.I. Carnerero\inst{3}
          \and
          M. Villata\inst{3}
          \and
          U. Bach\inst{5}
          \and
          P. Cassaro\inst{1}
          \and
          E. Liuzzo\inst{1}
          \and
          C. S. Buemi\inst{6}
          \and
          P. Leto\inst{6}
          \and
          C. Trigilio\inst{6}
          \and
          G. Umana\inst{6}
          \and    
          M. Bonato\inst{1}
          \and
          B. Patricelli\inst{7,8,9}
          \and
          A. Stamerra\inst{10}
          }

   \institute{INAF Istituto di Radioastronomia, Via Gobetti 101, I-40127 Bologna, Italy\\
              \email{nicola.marchili@inaf.it}
        \and
            INAF Istituto di Radioastronomia, Stazione di Medicina, Via Fiorentina 3513, I-40059, Villafontana (BO), Italy
        \and
            INAF, Osservatorio Astrofisico di Torino, via Osservatorio 20, I-10025 Pino
Torinese, Italy
        \and
            Dipartimento di Fisica e Astronomia “Augusto Righi”, Università di Bologna, via Gobetti 93/2, 40129 Bologna, Italy
        \and
            Max-Planck-Institut f{\"u}r Radioastronomie, Auf dem H{\"u}gel 69, D-53121 Bonn, Germany
        \and
            Osservatorio Astrofisico di Catania, INAF, Via S. Sofia 78, I-95123 Catania, Italy
        \and
            Physics Department, University of Pisa, Largo B. Pontecorvo 3, I-56127 Pisa, Italy
        \and
            INFN - Pisa, Largo B. Pontecorvo 3, I-56127 Pisa, Italy
        \and
            Scuola Normale Superiore di Pisa, P.zza dei Cavalieri 7, 56126 Pisa, Italy       
        \and
            INAF - Osservatorio Astronomico di Roma, Via Frascati 33, I-00078 Monte Porzio Catone (Rome), Italy
}

   \date{Received 28 July 2025; accepted 2 September 2025}

  \abstract
   {The extreme variability of blazars, in both timescale and amplitude, is generally explained as the effect of a relativistic jet closely aligned to the observer's line-of-sight. Due to causality arguments, variability characteristics translate into spatial information about the emitting region of blazars. Since radiation at different wavelengths
is emitted in different parts of the jet, multi-frequency observations provide us with a virtual view of the structure of the jet on different scales. Radio--$\gamma$-ray correlations, moreover, are essential to reveal where and how the high-energy radiation is produced.}
   {We present here the observations collected within the blazar radio monitoring program that we are running at the Medicina and Noto telescopes. It aims at investigating how the variability characteristics and spectral energy distribution of blazars evolve in time.}
   {Since 2004, observation have been performed at 5, 8, 24, and 43 GHz on 47 targets, with monthly cadence; the monitoring program is still active at frequencies of 8 and 24 GHz.}
   {The database we built in more than twenty years of activity comprises to date about 21000 flux density measurements. Some basic analysis tools have been applied to the data to characterise the detected variability and offer a first glance at the wealth of information that such a  program can provide about blazars.}
   {}

   \maketitle

\section{Introduction}

Blazars are among the most enigmatic and intriguing objects in the universe. They are a subclass of active galactic nuclei (AGN), characterised by an intense and highly variable emission across the whole electromagnetic spectrum. According to a well-established picture, at their cores lies a supermassive black hole, surrounded by an accretion disk of swirling gas and dust. The black hole powers a relativistic jet closely aligned to the observer's line-of-sight (\citealt{1979ApJ...232...34B}). Because of the bulk relativistic motion of the emitting plasma towards the observer, the apparent luminosity of blazars is affected by relativistic beaming, which can dramatically increase the amplitude and reduce the timescale of the observed flux density variations. Doppler boosting also results in apparent superluminal motion of the plasma itself. The blazar class is traditionally divided into two subclasses: Flat-Spectrum Radio Quasars (FSRQs) and BL Lacertae objects (BL Lacs). While both exhibit intense variability and possess relativistic jets, they differ in their spectral properties and the strength of their emission lines.

Variability studies play a pivotal role in unraveling the mysteries of blazars. They emit radiation across a broad range of wavelengths, from radio to gamma rays, with the radio regime offering unique insights into the nature and behaviour of these sources. The extreme variability of blazars challenges our understanding of the underlying mechanisms driving the intense emissions from their cores. Multi-wavelength variability studies, such the ones carried on by the Whole Earth Blazar Collaboration\footnote{\url{https://www.oato.inaf.it/blazars/webt/}} (WEBT), can provide invaluable information about the physical properties of the blazars’ emitting region (see, e.g., \citealt{2002A&A...390..407V,2006A&A...453..817V, 2017Natur.552..374R,2024A&A...692A..48R}). 
Born in 1997, the WEBT is an International Collaboration of astronomers whose aim is to monitor blazars in the radio, near-infrared and chiefly optical bands, to understand blazar variability on various time scales.
The insights provided by the variability characterisation is essential for discriminating among different emission models, estimating fundamental physical parameters (such as the size of the emitting regions, and Doppler boosting factors; see, e.g., \citealt{2017MNRAS.466.4625L}) and casting some light into the processes that power this class of AGNs. Since radiation at different wavelengths is emitted in different parts of the jet, multi-frequency observations provide us with a virtual view of the structure of the jet on different scales. Through the analysis of time delays between variations occurring at different wavelengths, it is possible to infer how perturbations propagate through the emitting regions located in different parts of the jet.

In the realm of radio observations, blazar variability is particularly intense, with fluctuations in radio flux densities occurring on timescales ranging from days to decades. Studying the time variability of blazars in the radio band sheds light on the dynamic processes occurring in their jets, providing clues about their formation, acceleration, and collimation. As confirmed by the F-GAMMA project (\citealt{2016A&A...596A..45F, 2019A&A...626A..60A}), long-term radio monitoring programs with high spectral coverage are of the utmost importance to disentangle the wide variety of variability manifestations observed in blazars. However, this goal can only be achieved by perpetuating monitoring programs for several decades, as recently shown by \cite{2025A&A...693A.318K}.

In 2004, a program for the monitoring of blazars, including both FSRQs and BL Lacs, started at the 32-m radio antennas of the INAF-Institute of Radioastronomy (IRA) in Medicina and Noto (see \citealt{2007A&A...464..175B}). 
A short-term blazar monitoring program had previously been run with the same telescopes from 1996 to 1999 \citep{2001A&A...379..755V}.
Observations were initially performed at frequencies of 5, 8, 22, and 43 GHz with a monthly cadence. Despite some discontinuities in frequency and time coverage (5 and 43 GHz observations have been dropped because of radio-frequency interference and problems with the receiver, respectively; periods of prolonged extraordinary maintenance of the antennas have created some gaps in the light curves), the ROBIN\footnote{\url{https://radio.oato.inaf.it/}} (Radio Observations of Blazars with INaf telescopes) program has been perpetuated across the years and is still operational, having reached 20 years of activity. The data collected within the program have been used in several studies and publications, mostly in correspondence of WEBT multi-frequency campaigns (see, e.g., \citealt{2025A&A...693A.196O, 2024A&A...692A..48R, 2006A&A...453..817V, 2007A&A...473..819R, 2021MNRAS.504.5629R, 2009A&A...507..769R, 2008A&A...492..389L, 2015MNRAS.450.2677C}) and in collaboration with the gamma-ray community (\citealt{2024MNRAS.529.3894M, 2023ApJS..266...37A, 2010Natur.463..919A,  2010ApJ...712..405V, 2012ApJ...754..114H, 2015A&A...576A.126A}).

In this paper, we offer an overview of the ROBIN program, providing the reader with basic information about the monitored sources and the observations. The collected data, which are stored in an online archive\footnote{\url{https://radio.oato.inaf.it/database/}}, will be made available upon request. We want to stress that an in-depth analysis of the variability of the sources is not within the purposes of this article, which aims instead to properly present and describe this important database. A thorough time series analysis of the light curves will be presented in a subsequent publication.

The paper is organised as follows: the source sample and the observations are briefly discussed in Sects. 2 and 3; the data reduction is described in Sect. 4, together with the results of a basic analysis of the data; in Sect. 5 we present our conclusions.

\section{The sources}
The database of the ROBIN program comprises flux density measurements of 47 sources, listed in Table \ref{table:1}; following the classification from \cite{2015Ap&SS.357...75M}, the sample includes 16  BL Lacs and 27 FSRQs, while for 4 objects no certain classification is given. For 0859+0454, the FSRQ classification is derived from \cite{2007ApJS..171...61H}. Currently, all the listed sources are still monitored, except 0323+342, which is no longer observed since the end of 2018, because its variability is comparable to the uncertainty in our flux density measurements, therefore it is impossible to derive reliable variability characteristics for it.

The sources were selected according to a few broad criteria: a high degree of variability at all the covered radio bands; observability from Medicina and Noto at reasonably high elevations; flux densities high enough to allow sufficiently precise measurements for a variability study. Some basic information about the sources is provided in Table \ref{table:1}; the redshift values have been extracted from the NASA/IPAC Extragalactic Database, NED\footnote{\url{https://ned.ipac.caltech.edu/}}.

\section{Observations}

Observations were almost entirely performed using the 32-m dishes located in Medicina and Noto, which are both owned and managed by INAF (National Institute for Astrophysics, Italy). In the first few months of the project, observations at 5, 8, and 22 GHz were obtained at both sites. Looking at the performances of the antennas, it was successively decided to split the observations, using the Medicina antenna for the monitoring at 8 and 22 GHz, and Noto for the 5 GHz one, since at this frequency the Medicina antenna was much more affected by radio frequency interference (RFI). In early 2006, a new 43 GHz receiver became available in Noto (see \citealt{2009A&A...507.1467L} for detailed information about observations performed using this receiver), allowing us to extend our monitoring to higher frequencies, where blazar variability is generally stronger.  One epoch of observation was performed at the Sardinia Radio Telescope, a 64-m dish equally managed by INAF, when both the 32-m dishes were unavailable.

Radio continuum acquisitions were carried out by means of On-The-Fly (OTF) cross-scans in Equatorial coordinates. Table \ref{t.telescopes} lists the telescope main features and employed configurations, while Table \ref{t.scans} provides the scanning parameters used during the observations. The number of cross-scans performed on a given source varied according to its expected flux density. Typical on-source integration times were 37.5\,s at 5~GHz, 40.0\,s at 8~GHz, 52.5\,s at 24 GHz, and 20.0\,s for 43 GHz acquisitions.   
Occasionally, some sources were observed in the same epoch and at the same frequency more than once. For the purposes of the present work, the flux density measurements from redundant observations have been averaged to provide a single measurement per epoch per frequency. Overall, the ROBIN database comprises 20938 datapoints, which translates, after the averaging, into the 12675 measurements that are presented in this paper.\\

Flux density calibration was carried out observing 3C\,123, 3C\,286, 3C\,295, NGC\,7027, whose reference flux densities were computed, for the observed band central frequency, according to \citet{2013ApJS..204...19P}. 
For 24-GHz observations, the atmospheric contribution was also taken into account in the calibration procedure; zenithal opacity was estimated by means of skydip acquisitions and then applied to calibrate the signal amplitude.

\begin{table*}
	\centering
	\caption{Telescope configurations and features, employed for the majority of the observations.}         
	\label{t.telescopes}    
	\begin{tabular}{l c c c c c}        
		\hline
		Telescope & Frequency range & Band  & Beamsize & $T_\mathrm{sys}^{(a)}$  & Max Gain \\ 
		          &    (GHz)  & & (arcmin) & (K)   & (K Jy$^{-1}$)   \\ 
		\hline                       
		Noto      & 4.65 - 5.02 & C & 7.5      & 30    & 0.14     \\
        Medicina  & 4.30 - 5.80 & C & 7.4      & 28    & 0.16     \\
		Noto      & 8.21 - 8.93 & X & 4.9      & 90    & 0.11     \\
        Medicina  & 8.18 - 8.86 & X & 4.9      & 38    & 0.14     \\
        Noto      & 22.0 - 24.0 & K & 1.7      & 110   & 0.08     \\
		Medicina  & 23.5 - 24.7 & K & 1.6      & 60    & 0.11     \\
        Noto      & 42.7 - 43.2 & Q & 0.9      & 100    & 0.08     \\
		\hline
	\end{tabular}\\
$^{(a)}$ $T_\mathrm{sys}$ pointing at Zenith, with $\tau_0=0.1$.\\
\end{table*}

\begin{table}
\centering                          
\caption{Cross-scan parameters.}         
\label{t.scans}    
\begin{tabular}{l c c c}        
\hline                 
Frequency   & Scan length & Scan speed  & Sampling \\ 
(GHz)  & (degrees)    & (arcmin s$^{-1}$)  & (s)      \\ 
\hline                       
5 & 1.0      & 4.0    & 0.040     \\
8 & 0.6      & 2.4    & 0.040     \\
24& 0.2      & 0.8    & 0.040     \\
43& 0.05     & 0.15   & 0.040     \\
\hline                                  
\end{tabular}
\end{table}

\section{Data reduction and analysis}
\label{sec:analysis} 

Our group developed an IDL pipeline (CAP, Cross-scan Analysis Pipeline) to reduce and analyse the cross-scans. It is fully described in \cite{2020MNRAS.492.2807G}. 
The pipeline performs the following operations: 

\begin{itemize}

    \item Data Rearrangement: the data set is organised by frequency and the purpose of the observed sources (calibrators, skydips, targets).

    \item Data Flagging: users can optionally use a GUI to inspect each subscan and assign flags, determining which data to include in the analysis. Since FITS files include both Left Circular Polarisation (LCP) and Right Circular Polarisation (RCP) data streams, which may be impacted differently by interferences (RFI), users can choose to include or exclude specific polarisations. 

    \item Skydips Reduction: for frequencies above 10 GHz, skydip acquisitions are used to estimate atmospheric opacity. These are fitted using atmospheric models, telescope parameters (e.g., receiver temperature), and weather data from the acquisition period, producing a table of zenithal opacity estimates.

    \item Calibrators Reduction: integrating acquisitions from flux density calibration sources involves aligning and integrating flagged subscans, correcting for antenna gain curve and atmospheric opacity as needed. Each subscan is handled individually due to elevation dependency. A Gaussian fitting is performed to measure signal amplitude in raw counts. The measured amplitude is corrected for pointing errors, and the theoretical flux density of the source is calculated based on the observed band, providing a counts-to-Jy conversion factor.

    \item Conversion Timeline: after processing data from flux density calibrators, users must decide how to interpolate the counts-to-Jy conversion factors. These factors, ideally constant after correcting for antenna and weather effects, may still fluctuate due to weather or instrument instabilities. Users can choose to average or linearly interpolate the conversion factors within a chosen time window.

    \item Targets Reduction: this process mirrors the "calibrators reduction" phase, where the raw amplitude from Gaussian fitting is multiplied by the appropriate calibration factor from the counts-to-Jy timeline. This results in a table of measured flux densities for each scan. LCP and RCP data streams are processed separately, producing independent flux density measurements. For each polarisation, the RA and Dec integrated semi-scans are averaged, yielding a single flux density. Partial results are documented in a separate table, and users can inspect the results of each original subscan if desired.

\end{itemize}

It must be noted that, as the operating and acquisition systems at the telescopes were still being developed during the many years of observations, the FITS file content details varied along time. CAP was conceived to manage this range of input formats.   

\subsection{System performance and statistics}
\label{sec:sysstat} 

At 24 GHz, even though the reduction pipeline is able to compensate flux density measurements for the pointing error, observations included pre-scan corrections of the antenna pointing. This was achieved by means of dedicated cross-scans on the targets themselves - when sufficiently bright - or on close bright sources. This allowed the subsequent acquisitions to be carried out with a pointing error mostly smaller than one tenth of the half-power beam width (HPBW). Errors larger than about 0.3 $\times$ HPBW were usually associated with bad weather conditions, also leading to poor quality data, so the associated acquisitions were discarded from the analysis. 
Individual measurements, especially in X band - the cleaner frequency band, less affected by RFI and weather effects - show flux density errors as low as $1\%$.     
To accommodate for flagging inaccuracies and calibration/weather fluctuations, we applied a conservative minimum value to the flux density errors: $3\%$ for C- and X-band measurements, $8\%$ for K-band flux densities, and $10-15\%$ for Q-band flux densities. Refined results are expected to be provided by a future version of the CAP pipeline, involving artificial intelligence procedures. 

\begin{figure}
	\includegraphics[width=1.09\columnwidth]{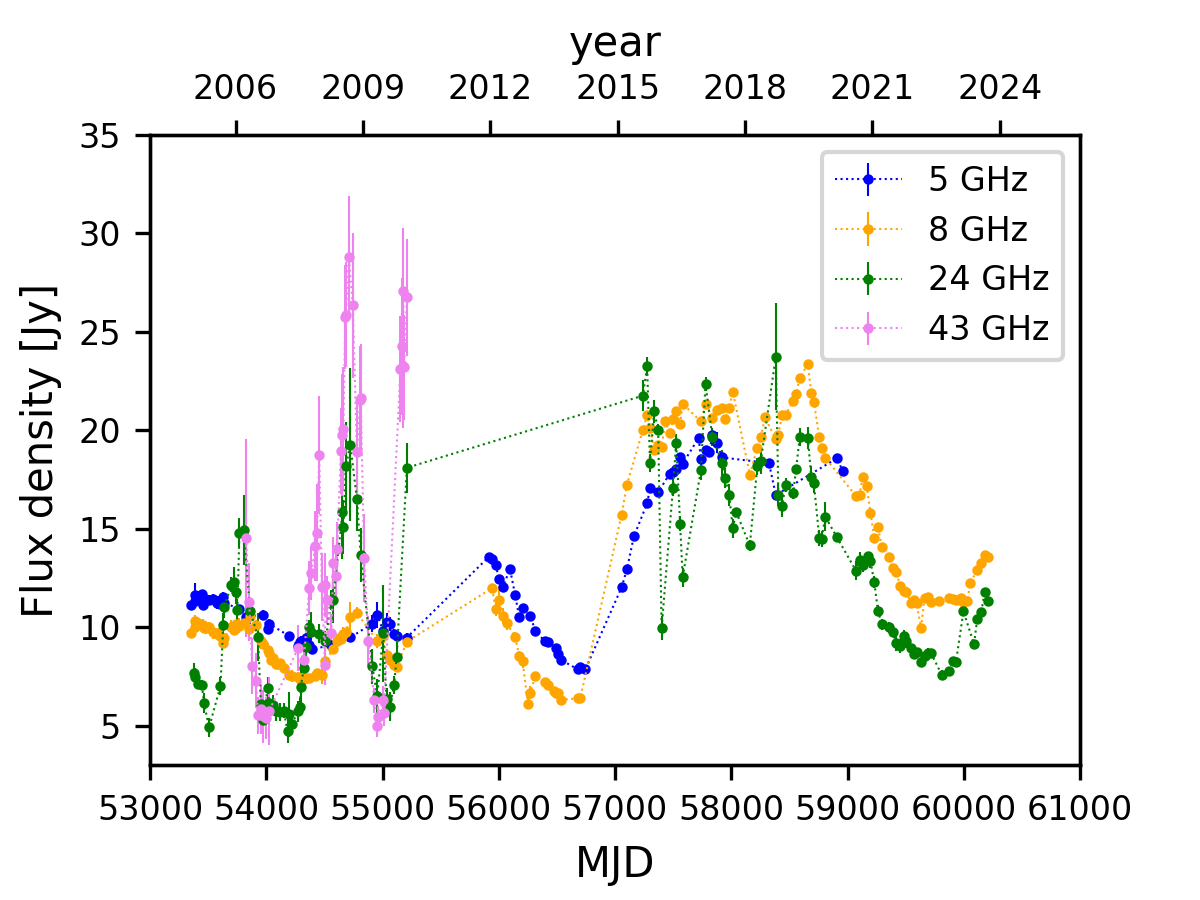}
    \caption{The light curves of 2251+158 (i.e. 3C454.3) from the ROBIN program. The 5, 8, 24, and 43 GHz data are plotted as blue, orange, green, and magenta dots, respectively.}
    \label{fig:lc}
\end{figure}

An example of the multi-frequency light curves collected within the ROBIN program is shown in Fig. \ref{fig:lc}.. The monthly cadence allows for adequate coverage of the fast variability of the sources: both the rising and falling part of the flares can be reconstructed with reasonable accuracy. The full set of the multi-frequency light curves is shown in the Appendix A.

\subsection{Basic variability characteristics}
\label{sec:dataChar}

Some simple analysis techniques have been applied to the data collected within the ROBIN program to extract basic information concerning their variability characteristics.

An estimation of the amplitude of the variability has been made by means of the intrinsic modulation index algorithm developed by \cite{2011ApJS..194...29R}:

\begin{equation}
    \overline{m} = \frac{\sigma_0}{\overline{S_0}},
\end{equation}
where $\sigma_0$ is the intrinsic standard deviation of the distribution of source flux densities in time, measured in units of the intrinsic source mean flux density, $\overline{S_0}$.

While the standard definition of the modulation index expresses the variability in terms of the standard deviation of the flux densities of the source, $\sigma$, and the average flux density, $\overline{S}$, the intrinsic modulation index uses the flux densities and variations as would be observed with uniform sampling of adequate cadence and zero observational error. To estimate these quantities, \cite{2011ApJS..194...29R} developed a likelihood analysis starting from the assumption that the true flux densities are normally distributed, and therefore both $\overline{S_0}$ and $\sigma_0$ are well defined, and can be inferred from the available data.

Despite the improvements introduced by the intrinsic modulation index with respect to the standard definition, some problems still persist. The assumption of stationarity, implicit in the definition of a true flux and a well-defined standard deviation, cannot be regarded as generally valid in the case of the radio emission of blazars. The radio variability of blazars often appears as a superposition of several components with different timescales: the longest timescales can be as long as decades (see e.g. \citealt{1992ApJ...396..469H}, and, more recently, \citealt{2025A&A...693A.318K}). This means that, despite the remarkable time span covered by the ROBIN program, the light curves obtained within the project are still not long enough to assume that $\overline{S_0}$ and $\sigma_0$ are always well defined. For this reason, the intrinsic modulation index values, which depend on $\overline{S_0}$ and $\sigma_0$, are sometimes almost as sensitive to the duration of the observations as the standard modulation index values, showing a tendency to increase with the length of the light curve.

Since the duration of the observations varies with the source and the observing frequency, the estimated $\overline{m}$ values, reported in Table \ref{table:2}, for the different bands, may not always be comparable to each other. An alternative estimation of the variability amplitude, not directly related to the temporal extension of the observations, has been achieved through a first-order structure function (\citealt{1985ApJ...296...46S}; henceforth, simply "structure function") that we calculate at a time lag of $\approx1.5$ years,

\begin{equation}
     SF_{1.5} = \frac{1}{N} \sum_{i,j} (S(t_i)-S(t_j))^2,
\end{equation}
where the sum extends to all N pairs ($t_i$, $t_j$) for which  1.4 yr $<$ ($t_i$ - $t_j$) $<$ 1.6 yr. 

$SF_{1.5}$ corresponds, roughly speaking, to twice the variance of the signal estimated using pairs of data with 1.5 years of separation. Starting from this consideration, we define a new function,  $SF_{1.5}^\prime$, as follows:

\begin{equation}
     SF_{1.5}^\prime = \frac{\sqrt{SF_{1.5}/2}}{\overline{S}}.
\end{equation}
Since this parameter, which is adimensional as the modulation index, provides a measure of the variability over a fixed interval of time, it is not affected by the dependence on the duration of the observations. The estimated $SF_{1.5}^\prime$ values for the different bands are reported in Table \ref{table:2}.

The time interval of 1.5 years we set for the $SF^\prime$ calculation was defined looking at \cite{2007A&A...469..899H}; by analysing decades-long light curves of a large sample of AGN at GHz frequencies, the authors found characteristics timescales (which, in case of multiple variability components, can be assumed to be the dominant ones) mostly between 1 and 2 years (see also \citealt{1992ApJ...396..469H}). A time interval of 1.5 years should therefore be sufficient to sample a large fraction of the dominant variability component for most of the sources. It should be noted that $SF_{1.5}^\prime$ does not take into account the uncertainty in the measurements. 

In Table \ref{table:2}, we also report the values of $SF_{1.5}^\prime/SF_{3.0}^\prime$, where $SF_{3.0}^\prime$ is calculated by summing the contribution of the pairs ($t_i$, $t_j$) for which  2.9 yr $<$ ($t_i$ - $t_j$) $<$ 3.1 yr. Since the structure function provides a way to evaluate the variability amplitude at different timescales, the ratio $SF_{1.5}^\prime/SF_{3.0}^\prime$ assesses how strong the variability on a short time scale is compared to the variability on longer timescales. In this sense, it can be considered as a rough indicator of the fastness of the flux density variations.

The average spectral index between 8 and 24 GHz, $\alpha$ (Table \ref{table:2}, Col. 14), were calculated over all the observing sessions for which both 8 and 24 GHz data are available. As a convention, we used $S(\nu) = S^\prime \nu^{\alpha}$, which means that positive $\alpha$ values denote an increase of flux density with frequency.

\begin{figure}
	\includegraphics[width=1.09\columnwidth]{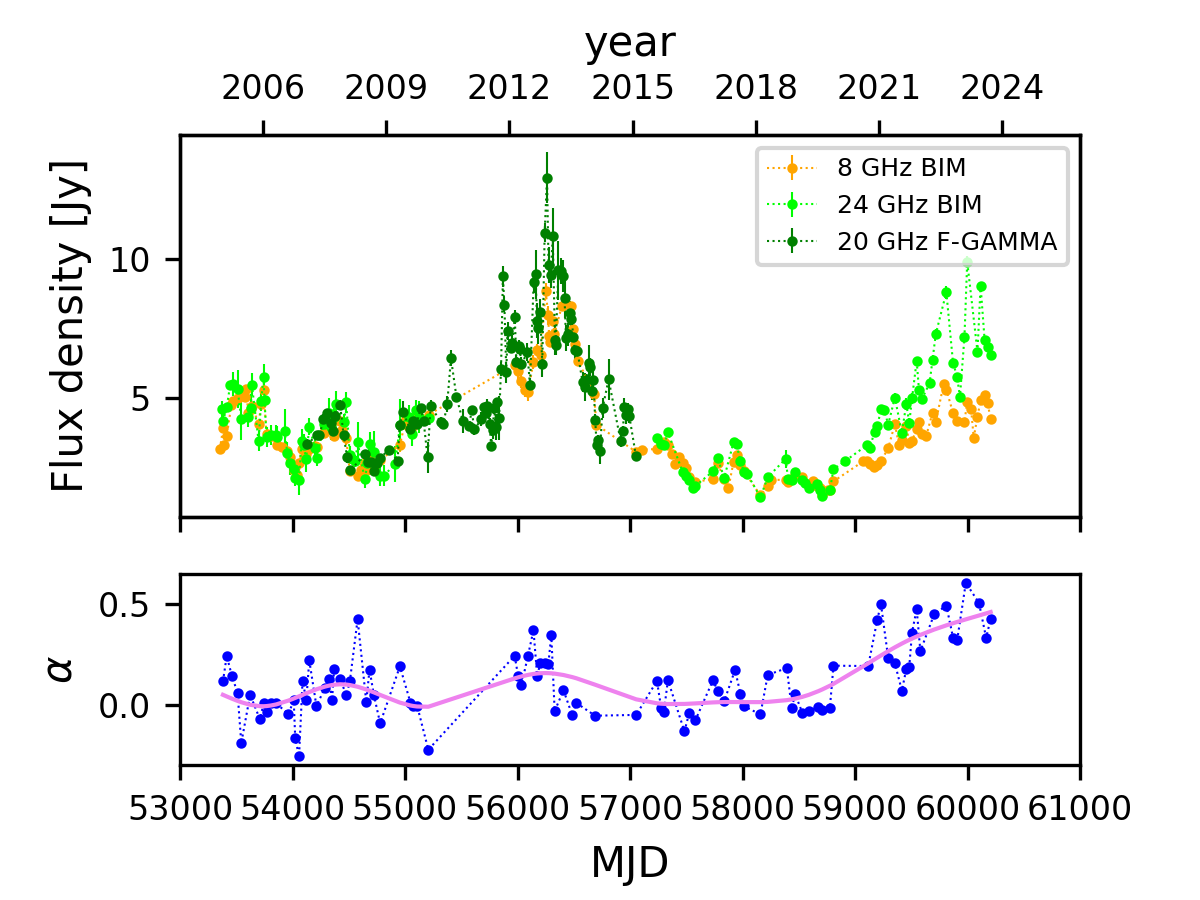}
    \caption{Upper panel: the ROBIN light curves of BL Lacertae at 24 and 8 GHz (light green and orange dots, respectively) are shown together with the re-scaled (see text for the details) 20 GHz ones from the F-GAMMA program (dark green dots). Lower panel: spectral indices calculated between the combined ROBIN and F-GAMMA data at 20-24 GHz and the ROBIN ones at 8 GHz. The clear increasing trend of the spectral index (magenta line) reflects the increasing discrepancy between the light curves in the last ~2000 days of observations.}
    \label{fig:bll_spind}
\end{figure}

An example of the spectral indices calculated for each epoch is shown in Fig \ref{fig:bll_spind} for BL Lacertae. Given the large gap in the 24 GHz, the data have been complemented with the 20 GHz observations from the F-GAMMA program (see \citealt{2016A&A...596A..45F} and \citealt{2019A&A...626A..60A}), re-scaled by a factor 1.05, which accounts for the slight difference in the flux densities measured at non-coincident frequencies and has been obtained by calculating the average ratio between quasi-simultaneous (i.e. separated by no more than 10 days) ROBIN and F-GAMMA flux density measurements. The spectral indices shown in the lower panel have been calculated after merging the ROBIN and the F-GAMMA 20-24 GHz data: the clear trend of an increasing spectral change between 24 and 8 GHz data is confirmed by the trend of $\alpha$, which may have started before MJD 58000 and seems about to reach a plateau.

\subsection{Comparison between variability amplitude estimators}
\label{sec:var_amp}

\begin{figure}
	\includegraphics[width=1.09\columnwidth]{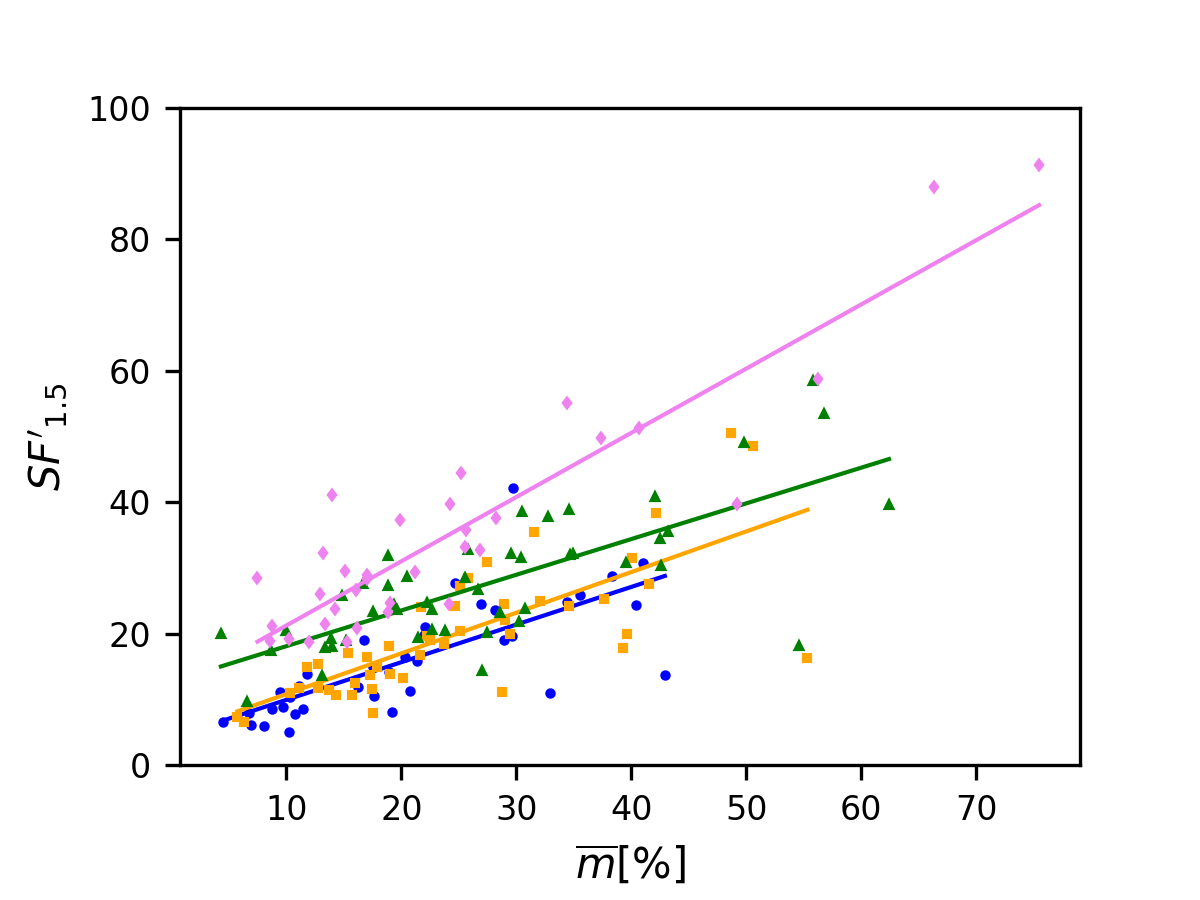}
    \caption{Comparison between the variability amplitude parameters, at 5 (blue dots), 8 (orange squares), 24 (green triangles), and 43 (magenta diamonds) GHz. The linear regression results are plotted as lines with the same colours.}
    \label{fig:lin_regr}
\end{figure}

A strong correlation is expected between $\overline{m}$ and $SF_{1.5}^\prime$. The results of our analysis, shown in Fig. \ref{fig:lin_regr} and Table \ref{tab:lin_reg}, confirm this expectation: the correlation coefficients calculated for the four observed bands range between 0.74 and 0.91. 

It should be noted that the 8 GHz light curve of 2344+514 has not been taken into account for the comparison between $\overline{m}$ and $SF_{1.5}^\prime$, because, according to $\overline{m}$, the source, at this frequency, does not show variability. Most likely the evaluation of this dataset as non-variable (which might be due to a combination of poor sampling and high uncertainties in the flux density measurements corresponding to possible flares) is not correct, as the variability of the source at 8 GHz has a timescale of months and is not consistent with white noise; the displayed variations, moreover, are correlated to the ones observed at 24 GHz, which are certainly source intrinsic.

Intuitively, we expect the correlation coefficient $r$ and the slope $B$ of the linear regression between $\overline{m}$ and $SF_{1.5}^\prime$ to increase with frequency; in fact, the higher the frequency, the stronger and faster should be the intrinsic variability of the source. As a consequence,  the $\overline{m}$ and the $SF_{1.5}^\prime$ values should tend to converge, and $B$ and $r$ should both tend to 1. In this regard, the behaviour of the variability parameters at 24 GHz is puzzling. The slope $B$ is considerably lower than at other frequencies, while the correlation coefficient aligns with the values estimated at 5 and 8 GHz, very far from the almost perfect correlation at 43 GHz. This behaviour seems to indicate that either $SF_{1.5}^\prime$ overestimates the variability (possibly because of a higher level of noise in the data at this frequency), or $\overline{m}$ underestimates it (this could be the consequence of an overestimation of the uncertainty in the data, which reduces estimated intrinsic variability).

The high correlation between  $\overline{m}$ and $SF_{1.5}^\prime$ at all frequencies shows that both tools are capable of reasonably well characterising the variability amplitude of the light curves. However, as we mentioned in Sect. \ref{sec:dataChar}, they both suffer from some bias. In the case of $SF_{1.5}^\prime$, it is important to take into account that  this estimator does not take into account the observational uncertainties, which causes an overestimation of the variability amplitude. Concerning $\overline{m}$, it does not distinguish variations according to the timescale on which they occur, resulting in a substantial dependence on the duration of the observations. By combining the two methods, however, we can reach a better understanding of the variability properties of the light curves.

\begin{table}[]
    \centering
        \caption{ Coefficients obtained for the linear regression between $\overline{m}$ and $SF_{1.5}^\prime$ at different frequencies.}
    \label{tab:lin_reg}
    \begin{tabular}{c c c c}
    \hline
      Frequency &   $r$ &  $A$ & $B$ \\
      (GHz) & & \\
      (1) & (2) & (3) & (4) \\
     \hline
      5   &  0.74 & 4.2 & 0.57 \\
      8   &  0.75 & 4.6 & 0.62 \\
     24   &  0.74 & 12.7 & 0.54 \\
     43   &  0.91 & 11.5 & 0.98 \\
     \hline
    \end{tabular}
    \tablefoot{ The Table reports: the frequency of the observations (Col. 1); the correlation coefficient $r$ (Col. 2); the intercept $A$ and the slope $B$ (Cols. 3 and 4, respectively).}
\end{table}

\subsection{Other parameters}

Unsurprisingly, there is a strong correlation between the spectral index of the sources and their variability amplitude; higher (i.e. more inverted) spectral indices correspond to more variable sources, independently of the estimator used for the variability amplitude and the frequency at which this is quantified. This is due to the fact that more inverted spectra are supposed to be associated with more compact emitting regions, in which flux density variations are more extreme. On average, the correlation of $\alpha$ with $SF_{1.5}^\prime$ provides higher correlation coefficients than the one between $\alpha$ and $\overline{m}$, but the difference is moderate (on average 0.51 for $SF_{1.5}^\prime$ and 0.45 for $\overline{m}$). The highest correlation, 0.56, is found with $SF_{1.5}^\prime$(24 GHz), while the correlation with $\overline{m}$(24 GHz) is only 0.44. This is an interesting indication with respect to the issue discussed in Sect. \ref{sec:var_amp} concerning the linear regression coefficients $B$ and $r$ at 24 GHz. The higher correlation between $\alpha$ and $SF_{1.5}^\prime$ suggests that, in this case, the latter provides us with a more faithful evaluation of the variability properties compared to $\overline{m}$.

No correlation is found between $\overline{m}$ and $SF_{1.5}^\prime/SF_{3.0}^\prime$ at any observing frequency, which suggests that higher variability amplitudes are not systematically associated to an increase or a decrease of the variability timescale. Also between $\alpha$ and $SF_{1.5}^\prime/SF_{3.0}^\prime$ no obvious correlation is seen, which, again, can be interpreted as evidence that the spectral index is independent of the fastness of the variations.

\begin{figure}
	\includegraphics[width=1.09\columnwidth]{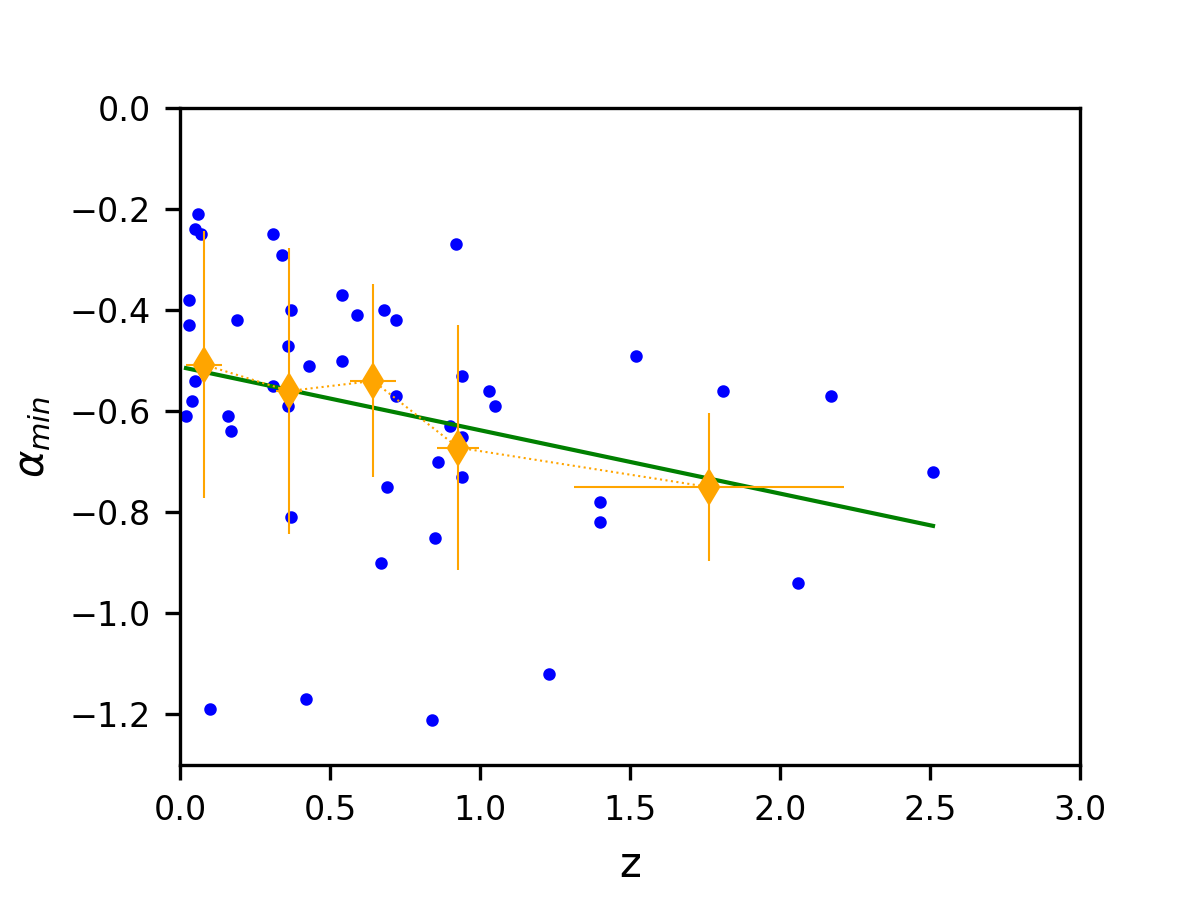}
    \caption{Minimum value of the spectral index calculated over all the available epochs is plotted versus redshift. Both a linear regression of the data (green line) and the average calculated over increasingly larger bins (in order to keep the number of contributing points comparable; orange diamonds) show a moderate anti-correlation between the two parameters.}
    \label{fig:si_z}
\end{figure}

All the parameters reported in Table \ref{table:2} have been checked for possible correlations with redshift. None was found; however the minimum values of the spectral index calculated over all epochs show a weak anti-correlation with redshift at a level of --0.30 (see Fig.\ref{fig:si_z}), which implies a probability of by chance correlation of $\sim4\%$. In particular, while at low redshift the minimum spectral index can go from very steep (--1.19) to almost flat (--0.21), with an average value of  about $-0.5$, at high redshift it does not reach higher values than --0.56. Given the low number of datapoints, this result should be considered cautiously.

\subsection{BL Lacs vs FSRQs}

\begin{figure}
	\includegraphics[width=1.09\columnwidth]{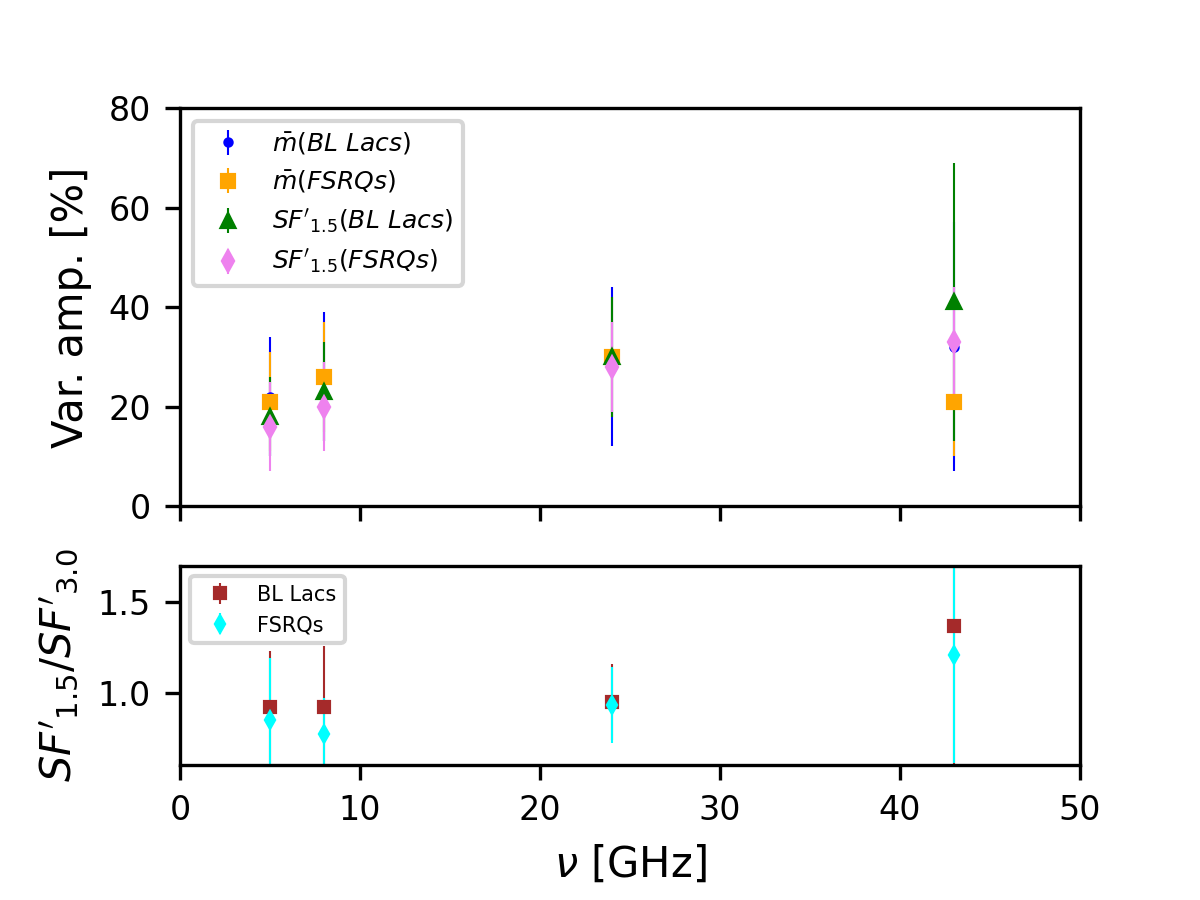}
    \caption{The mean variability parameters for BL Lacs and FSRQs are plotted separately as a function of frequency. On the upper panel are shown the $\overline{m}$ and $SF_{1.5}^\prime$ values, while on the lower one the fastness of the variability calculated through $SF_{1.5}^\prime/SF_{3.0}^\prime$.}
    \label{fig:bll_fsrq}
\end{figure}

The mean variability parameters for BL Lacs and FSRQs are plotted separately on Fig. \ref{fig:bll_fsrq}. Given the relatively low number of objects in the sample and the large spread in their characteristics, the uncertainties in the mean values are large, and the results discussed below should be considered as purely indicative.

Both $\overline{m}$ and $SF_{1.5}^\prime$ show that the variability of our sample of BL Lacs is higher than that of FSRQs; the difference, however, is remarkable only at 43 GHz. This result is in agreement with the one documented by \cite{2017MNRAS.467.4565L}; a higher variability for FSRQs, concerning the first 2.5 years of monitoring of the F-GAMMA program, is reported by \cite{2016A&A...596A..45F}, which though estimates the variability amplitude in terms of flux density variance instead of modulation index.

The light curves show a general increase of the variability with frequency, as expected from previous studies (see, e.g., \citealt{2016A&A...596A..45F}), partially contradicted by the sudden drop of $\overline{m}$(43 GHz) for FSRQs. This drop is not seen in $SF_{1.5}^\prime$(43 GHz), which may be surprising given the high level of correlations between the two parameters at this frequency (see discussion in Sect. \ref{sec:var_amp}). The difference between the two is illustrated by the high value of the intercept $A$ (see Table \ref{tab:lin_reg}), which reflects a systematically higher value of $SF_{1.5}^\prime$ with respect to $\overline{m}$; this seems a further confirmation of the tendency of $\overline{m}$ to underestimate the intrinsic variability of the sources.

For the mean values of $SF_{1.5}^\prime/SF_{3.0}^\prime$ we see systematically higher values for BL Lacs than for FSRQs, with a tendency to increase with frequency, suggesting that the former show generally faster variability than the latter, and that variability timescales decrease with frequency. Note that values higher than 1, as obtained at 43 GHz, do not have an obvious physical interpretation, as $SF_{1.5}^\prime/SF_{3.0}^\prime$ should oscillate around 1 when the variability timescale is lower than 1.5  years, unless the light curves display quasi-periodic oscillations. The behaviour here observed is certainly due to the large uncertainty in the data. Nevertheless the increasing trend of fastness with frequency is consistent with our expectations.

 We evaluated the presence of statistical difference in the variability parameters of BL Lacs and FSRQs through a Kolmogorov-Smirnov test; no significant difference was found.

\section{Summary}

We presented the data collected within the ROBIN program. Started in 2004, the project has so far gathered about 21000 flux density measurements at frequencies of 5, 8, 24, and 43 GHz for a sample of 47 blazars. The light curves of all the sources in the sample, which comprises 27 FSRQs and 16 BL Lacs, plus 4 objects with uncertain classification, are shown in the  Appendix A, where the four frequencies, in increasing order, are plotted as blue, orange, green, and magenta dots.

We applied some basic analysis tools to the data, for the estimation of the variability amplitude (through both the intrinsic modulation index and a structure-function-based algorithm), the fastness of the variability, and the spectral index between 8 and 24 GHz. A linear regression has been applied to the estimated parameters in a search for correlations among them. A significant correlation has been found only between spectral index and variability amplitude of the sources, independently of the way the latter is estimated: the most variable sources are the ones with the most inverted spectrum. 

No difference is found between the variability parameters of BL Lacs and FSRQs on a significant statistical level, estimated through a Kolmogorov-Smirnov test. In general, the former are more variable at all wavelengths than the latter; also, their variability occurs on faster timescales and their spectral index is higher, but the differences are  not significant.

After reaching and passing twenty years of activity, the ROBIN program has recently been renewed for five more years,  which will corroborate an archive that is among the most long-lived in this field of research. The datasets of single sources from the ROBIN program are available from the corresponding author upon reasonable request.
 
 \section{Data availability}
 
 Tables \ref{table:1} and \ref{table:2} are only available in electronic form at the CDS via anonymous ftp to cdsarc.u-strasbg.fr (130.79.128.5) or via http://cdsweb.u-strasbg.fr/cgi-bin/qcat?J/A+A/.

\begin{acknowledgements}
The Medicina and Noto radio telescopes are funded by the Ministry of University and Research (MUR) and are operated as National Facilities by the National Institute for Astrophysics (INAF).
The Sardinia Radio Telescope is funded by the Ministry of University and Research (MUR), Italian Space Agency (ASI), and the Autonomous Region of Sardinia (RAS) and is operated as National Facility by the National Institute for Astrophysics (INAF).
We acknowledge financial support from the INAF Fundamental
Research Funding Call 2023 (PI: Raiteri).

\end{acknowledgements}

\bibliographystyle{aa}
\bibliography{bibliography}

\begin{thebibliography}{33}
\expandafter\ifx\csname natexlab\endcsname\relax\def\natexlab#1{#1}\fi

\bibitem[{{Abdo} {et~al.}(2010){Abdo}, {Ackermann}, {Ajello}, {Axelsson},
  {Baldini}, {Ballet}, {Barbiellini}, {Bastieri}, {Baughman}, {Bechtol},
  {Bellazzini}, {Berenji}, {Blandford}, {Bloom}, {Bock}, {Bogart}, {Bonamente},
  {Borgland}, {Bouvier}, {Bregeon}, {Brez}, {Brigida}, {Bruel}, {Burnett},
  {Buson}, {Caliandro}, {Cameron}, {Caraveo}, {Casandjian}, {Cavazzuti},
  {Cecchi}, {{\c{C}}elik}, {Chekhtman}, {Cheung}, {Chiang}, {Ciprini}, {Claus},
  {Cohen-Tanugi}, {Collmar}, {Cominsky}, {Conrad}, {Corbel}, {Corbet},
  {Costamante}, {Cutini}, {Dermer}, {de Angelis}, {de Palma}, {Digel}, {Do
  Couto E Silva}, {Drell}, {Dubois}, {Dumora}, {Farnier}, {Favuzzi}, {Fegan},
  {Ferrara}, {Focke}, {Fortin}, {Frailis}, {Fuhrmann}, {Fukazawa}, {Funk},
  {Fusco}, {Gargano}, {Gasparrini}, {Gehrels}, {Germani}, {Giebels},
  {Giglietto}, {Giommi}, {Giordano}, {Giroletti}, {Glanzman}, {Godfrey},
  {Grenier}, {Grove}, {Guillemot}, {Guiriec}, {Hanabata}, {Harding},
  {Hayashida}, {Hays}, {Horan}, {Hughes}, {Iafrate}, {Itoh}, {Jackson},
  {J{\'o}hannesson}, {Johnson}, {Johnson}, {Kadler}, {Kamae}, {Katagiri},
  {Kataoka}, {Kawai}, {Kerr}, {Kn{\"o}dlseder}, {Kocian}, {Kuss}, {Lande},
  {Larsson}, {Latronico}, {Lemoine-Goumard}, {Longo}, {Loparco}, {Lott},
  {Lovellette}, {Lubrano}, {Macquart}, {Madejski}, {Makeev}, {Max-Moerbeck},
  {Mazziotta}, {McConville}, {McEnery}, {McGlynn}, {Meurer}, {Michelson},
  {Mitthumsiri}, {Mizuno}, {Moiseev}, {Monte}, {Monzani}, {Morselli},
  {Moskalenko}, {Murgia}, {Nestoras}, {Nolan}, {Norris}, {Nuss}, {Ohsugi},
  {Okumura}, {Omodei}, {Orlando}, {Ormes}, {Paneque}, {Panetta}, {Parent},
  {Pavlidou}, {Pearson}, {Pelassa}, {Pepe}, {Pesce-Rollins}, {Piron}, {Porter},
  {Rain{\`o}}, {Rando}, {Razzano}, {Readhead}, {Reimer}, {Reimer}, {Reposeur},
  {Reyes}, {Richards}, {Rochester}, {Rodriguez}, {Roth}, {Ryde}, {Sadrozinski},
  {Sanchez}, {Sander}, {Saz Parkinson}, {Scargle}, {Sgr{\`o}}, {Shaw},
  {Shrader}, {Siskind}, {Smith}, {Smith}, {Spandre}, {Spinelli}, {Stawarz},
  {Stevenson}, {Strickman}, {Suson}, {Tajima}, {Takahashi}, {Takahashi},
  {Tanaka}, {Taylor}, {Thayer}, {Thayer}, {Thompson}, {Tibaldo}, {Torres},
  {Tosti}, {Tramacere}, {Uchiyama}, {Usher}, {Vasileiou}, {Vilchez}, {Vitale},
  {Waite}, {Wang}, {Wehrle}, {Winer}, {Wood}, {Ylinen}, {Zensus}, {Uemura},
  {Ikejiri}, {Kawabata}, {Kino}, {Sakimoto}, {Sasada}, {Sato}, {Yamanaka},
  {Villata}, {Raiteri}, {Agudo}, {Aller}, {Aller}, {Angelakis}, {Arkharov},
  {Bach}, {Ben{\'\i}tez}, {Berdyugin}, {Blinov}, {Boettcher}, {Buemi}, {Chen},
  {Dolci}, {Dultzin}, {Efimova}, {Gurwell}, {Gusbar}, {G{\'o}mez}, {Heidt},
  {Hiriart}, {Hovatta}, {Jorstad}, {Konstantinova}, {Kopatskaya}, {Koptelova},
  {Kurtanidze}, {Lahteenmaki}, {Larionov}, {Larionova}, {Leto}, {Lin},
  {Lindfors}, {Marscher}, {McHardy}, {Melnichuk}, {Mommert}, {Nilsson}, {di
  Paola}, {Reinthal}, {Richter}, {Roca-Sogorb}, {Roustazadeh}, {Sigua},
  {Takalo}, {Tornikoski}, {Trigilio}, {Troitsky}, {Umana}, {Villforth},
  {Grainge}, {Moderski}, {Nalewajko}, {Sikora}, {Fermi LAT Collaboration}, \&
  {Members of the 3C Multi-Band Campaign}}]{2010Natur.463..919A}
{Abdo}, A.~A., {Ackermann}, M., {Ajello}, M., {et~al.} 2010, \nat, 463, 919

\bibitem[{{Abe} {et~al.}(2023){Abe}, {Abe}, {Acciari}, {Agudo}, {Aniello},
  {Ansoldi}, {Antonelli}, {Arbet-Engels}, {Arcaro}, {Artero}, {Asano}, {Baack},
  {Babi{\'c}}, {Baquero}, {Barres de Almeida}, {Barrio}, {Batkovi{\'c}},
  {Baxter}, {Becerra Gonz{\'a}lez}, {Bednarek}, {Bernardini}, {Bernardos},
  {Berti}, {Besenrieder}, {Bhattacharyya}, {Bigongiari}, {Biland}, {Blanch},
  {Bonnoli}, {Bo{\v{s}}njak}, {Burelli}, {Busetto}, {Carosi},
  {Carretero-Castrillo}, {Castro-Tirado}, {Ceribella}, {Chai}, {Chilingarian},
  {Cikota}, {Colombo}, {Contreras}, {Cortina}, {Covino}, {D'Amico}, {D'Elia},
  {da Vela}, {Dazzi}, {de Angelis}, {de Lotto}, {Del Popolo}, {Delfino},
  {Delgado}, {Delgado Mendez}, {Depaoli}, {di Pierro}, {di Venere}, {Do Souto
  Espi{\~n}eira}, {Dominis Prester}, {Donini}, {Dorner}, {Doro}, {Elsaesser},
  {Emery}, {Escudero}, {Fallah Ramazani}, {Fari{\~n}a}, {Fattorini}, {Foffano},
  {Font}, {Fruck}, {Fukami}, {Fukazawa}, {Garc{\'\i}a L{\'o}pez},
  {Garczarczyk}, {Gasparyan}, {Gaug}, {Giesbrecht Paiva}, {Giglietto},
  {Giordano}, {Gliwny}, {Godinovi{\'c}}, {Grau}, {Green}, {Green}, {Hadasch},
  {Hahn}, {Hassan}, {Heckmann}, {Herrera}, {Hrupec}, {H{\"u}tten}, {Imazawa},
  {Inada}, {Iotov}, {Ishio}, {Jim{\'e}nez Mart{\'\i}nez}, {Jormanainen},
  {Kerszberg}, {Kobayashi}, {Kubo}, {Kushida}, {Lamastra}, {Lelas}, {Leone},
  {Lindfors}, {Linhoff}, {Lombardi}, {Longo}, {L{\'o}pez-Coto},
  {L{\'o}pez-Moya}, {L{\'o}pez-Oramas}, {Loporchio}, {Lorini}, {Lyard},
  {Machado de Oliveira Fraga}, {Majumdar}, {Makariev}, {Maneva}, {Mang},
  {Manganaro}, {Mangano}, {Mannheim}, {Mariotti}, {Mart{\'\i}nez},
  {Mas-Aguilar}, {Mazin}, {Menchiari}, {Mender}, {Mi{\'c}anovi{\'c}}, {Miceli},
  {Miener}, {Miranda}, {Mirzoyan}, {Molina}, {Mondal}, {Moralejo}, {Morcuende},
  {Moreno}, {Nakamori}, {Nanci}, {Nava}, {Neustroev}, {Nievas Rosillo},
  {Nigro}, {Nilsson}, {Nishijima}, {Njoh Ekoume}, {Noda}, {Nozaki}, {Ohtani},
  {Oka}, {Okumura}, {Otero-Santos}, {Paiano}, {Palatiello}, {Paneque},
  {Paoletti}, {Paredes}, {Pavleti{\'c}}, {Persic}, {Pihet}, {Pirola},
  {Podobnik}, {Prada Moroni}, {Prandini}, {Principe}, {Priyadarshi}, {Rhode},
  {Rib{\'o}}, {Rico}, {Righi}, {Rugliancich}, {Sahakyan}, {Saito}, {Sakurai},
  {Satalecka}, {Saturni}, {Schleicher}, {Schmidt}, {Schmuckermaier},
  {Schubert}, {Schweizer}, {Sitarek}, {Sliusar}, {Sobczynska}, {Spolon},
  {Stamerra}, {Stri{\v{s}}kovi{\'c}}, {Strom}, {Strzys}, {Suda}, {Suri{\'c}},
  {Tajima}, {Takahashi}, {Takeishi}, {Tavecchio}, {Temnikov}, {Terauchi},
  {Terzi{\'c}}, \& {Teshima}}]{2023ApJS..266...37A}
{Abe}, H., {Abe}, S., {Acciari}, V.~A., {et~al.} 2023, \apjs, 266, 37

\bibitem[{{Aleksi{\'c}} {et~al.}(2015){Aleksi{\'c}}, {Ansoldi}, {Antonelli},
  {Antoranz}, {Babic}, {Bangale}, {Barres de Almeida}, {Barrio}, {Becerra
  Gonz{\'a}lez}, {Bednarek}, {Berger}, {Bernardini}, {Biland}, {Blanch},
  {Bock}, {Bonnefoy}, {Bonnoli}, {Borracci}, {Bretz}, {Carmona}, {Carosi},
  {Carreto Fidalgo}, {Colin}, {Colombo}, {Contreras}, {Cortina}, {Covino}, {Da
  Vela}, {Dazzi}, {De Angelis}, {De Caneva}, {De Lotto}, {Delgado Mendez},
  {Doert}, {Dom{\'\i}nguez}, {Dominis Prester}, {Dorner}, {Doro}, {Einecke},
  {Eisenacher}, {Elsaesser}, {Farina}, {Ferenc}, {Fonseca}, {Font}, {Frantzen},
  {Fruck}, {Garc{\'\i}a L{\'o}pez}, {Garczarczyk}, {Garrido Terrats}, {Gaug},
  {Giavitto}, {Godinovi{\'c}}, {Gonz{\'a}lez Mu{\~n}oz}, {Gozzini}, {Hadamek},
  {Hadasch}, {Herrero}, {Hildebrand}, {Hose}, {Hrupec}, {Idec}, {Kadenius},
  {Kellermann}, {Knoetig}, {Krause}, {Kushida}, {La Barbera}, {Lelas},
  {Lewandowska}, {Lindfors}, {Longo}, {Lombardi}, {L{\'o}pez},
  {L{\'o}pez-Coto}, {L{\'o}pez-Oramas}, {Lorenz}, {Lozano}, {Makariev},
  {Mallot}, {Maneva}, {Mankuzhiyil}, {Mannheim}, {Maraschi}, {Marcote},
  {Mariotti}, {Mart{\'\i}nez}, {Mazin}, {Menzel}, {Meucci}, {Miranda},
  {Mirzoyan}, {Moralejo}, {Munar-Adrover}, {Nakajima}, {Niedzwiecki},
  {Nilsson}, {Nowak}, {Orito}, {Overkemping}, {Paiano}, {Palatiello},
  {Paneque}, {Paoletti}, {Paredes}, {Paredes-Fortuny}, {Partini}, {Persic},
  {Prada}, {Prada Moroni}, {Prandini}, {Preziuso}, {Puljak}, {Reinthal},
  {Rhode}, {Rib{\'o}}, {Rico}, {RodriguezGarcia}, {R{\"u}gamer}, {Saggion},
  {Saito}, {Salvati}, {Satalecka}, {Scalzotto}, {Scapin}, {Schultz},
  {Schweizer}, {Shore}, {Sillanp{\"a}{\"a}}, {Sitarek}, {Snidaric},
  {Sobczynska}, {Spanier}, {Stamatescu}, {Stamerra}, {Steinbring}, {Storz},
  {Sun}, {Suri{\'c}}, {Takalo}, {Tavecchio}, {Temnikov}, {Terzi{\'c}},
  {Tescaro}, {Teshima}, {Thaele}, {Tibolla}, {Torres}, {Toyama}, {Treves},
  {Uellenbeck}, {Vogler}, {Wagner}, {Zandanel}, {Zanin}, {MAGIC Collaboration},
  {Archambault}, {Behera}, {Beilicke}, {Benbow}, {Bird}, {Buckley}, {Bugaev},
  {Cerruti}, {Chen}, {Ciupik}, {Collins-Hughes}, {Cui}, {Dumm}, {Eisch},
  {Falcone}, {Federici}, {Feng}, {Finley}, {Fleischhack}, {Fortin}, {Fortson},
  {Furniss}, {Griffin}, {Griffiths}, {Grube}, {Gyuk}, {Hanna}, {Holder},
  {Hughes}, {Humensky}, {Johnson}, {Kaaret}, {Kertzman}, {Khassen}, {Kieda},
  {Krawczynski}, {Krennrich}, {Kumar}, {Lang}, {Maier}, {McArthur}, {Meagher},
  {Moriarty}, {Mukherjee}, {Ong}, {Otte}, {Park}, {Pichel}, {Pohl}, {Popkow},
  {Prokoph}, {Quinn}, {Ragan}, {Rajotte}, {Reynolds}, {Richards}, {Roache},
  {Rovero}, {Sembroski}, {Shahinyan}, {Staszak}, {Telezhinsky}, {Theiling},
  {Tucci}, {Tyler}, {Varlotta}, {Wakely}, {Weekes}, {Weinstein}, {Welsing},
  {Wilhelm}, {Williams}, {Zitzer}, {VERITAS Collaboration}, {Villata},
  {Raiteri}, {Aller}, {Aller}, {Chen}, {Jordan}, {Koptelova}, {Kurtanidze},
  {L{\"a}hteenm{\"a}ki}, {McBreen}, {Larionov}, {Lin}, {Nikolashvili},
  {Angelakis}, {Capalbi}, {Carrami{\~n}ana}, {Carrasco}, {Cassaro}, {Cesarini},
  {Fuhrmann}, {Giroletti}, {Hovatta}, {Krichbaum}, {Krimm}, {Max-Moerbeck},
  {Moody}, {Maccaferri}, {Mori}, {Nestoras}, {Orlati}, {Pace}, {Pearson},
  {Perri}, {Readhead}, {Richards}, {Sadun}, {Sakamoto}, {Tammi}, {Tornikoski},
  {Yatsu}, \& {Zook}}]{2015A&A...576A.126A}
{Aleksi{\'c}}, J., {Ansoldi}, S., {Antonelli}, L.~A., {et~al.} 2015, \aap, 576,
  A126

\bibitem[{{Angelakis} {et~al.}(2019){Angelakis}, {Fuhrmann}, {Myserlis},
  {Zensus}, {Nestoras}, {Karamanavis}, {Marchili}, {Krichbaum}, {Kraus}, \&
  {Rachen}}]{2019A&A...626A..60A}
{Angelakis}, E., {Fuhrmann}, L., {Myserlis}, I., {et~al.} 2019, \aap, 626, A60

\bibitem[{{Bach} {et~al.}(2007){Bach}, {Raiteri}, {Villata}, {Fuhrmann},
  {Buemi}, {Larionov}, {Letog}, {Arkharov}, {Coloma}, {di Paola}, {Dolci},
  {Efimova}, {Forn{\'e}}, {Ibrahimov}, {Hagen-Thorn}, {Konstantinova},
  {Kopatskaya}, {Lanteri}, {Kurtanidze}, {Maccaferri}, {Nikolashvili},
  {Orlati}, {Ros}, {Tosti}, {Trigilio}, \& {Umana}}]{2007A&A...464..175B}
{Bach}, U., {Raiteri}, C.~M., {Villata}, M., {et~al.} 2007, \aap, 464, 175

\bibitem[{{Blandford} \& {K{\"o}nigl}(1979)}]{1979ApJ...232...34B}
{Blandford}, R.~D. \& {K{\"o}nigl}, A. 1979, \apj, 232, 34

\bibitem[{{Carnerero} {et~al.}(2015){Carnerero}, {Raiteri}, {Villata},
  {Acosta-Pulido}, {D'Ammando}, {Smith}, {Larionov}, {Agudo}, {Ar{\'e}valo},
  {Arkharov}, {Bach}, {Bachev}, {Ben{\'\i}tez}, {Blinov}, {Bozhilov}, {Buemi},
  {Bueno Bueno}, {Carosati}, {Casadio}, {Chen}, {Damljanovic}, {di Paola},
  {Efimova}, {Ehgamberdiev}, {Giroletti}, {G{\'o}mez}, {Gonz{\'a}lez-Morales},
  {Grinon-Marin}, {Grishina}, {Gurwell}, {Hiriart}, {Hsiao}, {Ibryamov},
  {Jorstad}, {Joshi}, {Kopatskaya}, {Kurtanidze}, {Kurtanidze},
  {L{\"a}hteenm{\"a}ki}, {Larionova}, {Larionova}, {L{\'a}zaro}, {Leto}, {Lin},
  {Lin}, {Manilla-Robles}, {Marscher}, {McHardy}, {Metodieva}, {Mirzaqulov},
  {Mokrushina}, {Molina}, {Morozova}, {Nikolashvili}, {Orienti}, {Ovcharov},
  {Panwar}, {Pastor Yabar}, {Puerto Gim{\'e}nez}, {Ramakrishnan}, {Richter},
  {Rossini}, {Sigua}, {Strigachev}, {Taylor}, {Tornikoski}, {Trigilio},
  {Troitskaya}, {Troitsky}, {Umana}, {Valcheva}, {Velasco}, {Vince}, {Wehrle},
  \& {Wiesemeyer}}]{2015MNRAS.450.2677C}
{Carnerero}, M.~I., {Raiteri}, C.~M., {Villata}, M., {et~al.} 2015, \mnras,
  450, 2677

\bibitem[{{Fuhrmann} {et~al.}(2016){Fuhrmann}, {Angelakis}, {Zensus},
  {Nestoras}, {Marchili}, {Pavlidou}, {Karamanavis}, {Ungerechts}, {Krichbaum},
  {Larsson}, {Lee}, {Max-Moerbeck}, {Myserlis}, {Pearson}, {Readhead},
  {Richards}, {Sievers}, \& {Sohn}}]{2016A&A...596A..45F}
{Fuhrmann}, L., {Angelakis}, E., {Zensus}, J.~A., {et~al.} 2016, \aap, 596, A45

\bibitem[{{Giroletti} \& {Righini}(2020)}]{2020MNRAS.492.2807G}
{Giroletti}, M. \& {Righini}, S. 2020, \mnras, 492, 2807

\bibitem[{{Hayashida} {et~al.}(2012){Hayashida}, {Madejski}, {Nalewajko},
  {Sikora}, {Wehrle}, {Ogle}, {Collmar}, {Larsson}, {Fukazawa}, {Itoh},
  {Chiang}, {Stawarz}, {Blandford}, {Richards}, {Max-Moerbeck}, {Readhead},
  {Buehler}, {Cavazzuti}, {Ciprini}, {Gehrels}, {Reimer}, {Szostek}, {Tanaka},
  {Tosti}, {Uchiyama}, {Kawabata}, {Kino}, {Sakimoto}, {Sasada}, {Sato},
  {Uemura}, {Yamanaka}, {Greiner}, {Kruehler}, {Rossi}, {Macquart}, {Bock},
  {Villata}, {Raiteri}, {Agudo}, {Aller}, {Aller}, {Arkharov}, {Bach},
  {Ben{\'\i}tez}, {Berdyugin}, {Blinov}, {Blumenthal}, {B{\"o}ttcher}, {Buemi},
  {Carosati}, {Chen}, {Di Paola}, {Dolci}, {Efimova}, {Forn{\'e}}, {G{\'o}mez},
  {Gurwell}, {Heidt}, {Hiriart}, {Jordan}, {Jorstad}, {Joshi}, {Kimeridze},
  {Konstantinova}, {Kopatskaya}, {Koptelova}, {Kurtanidze},
  {L{\"a}hteenm{\"a}ki}, {Lamerato}, {Larionov}, {Larionova}, {Larionova},
  {Leto}, {Lindfors}, {Marscher}, {McHardy}, {Molina}, {Morozova},
  {Nikolashvili}, {Nilsson}, {Reinthal}, {Roustazadeh}, {Sakamoto}, {Sigua},
  {Sillanp{\"a}{\"a}}, {Takalo}, {Tammi}, {Taylor}, {Tornikoski}, {Trigilio},
  {Troitsky}, \& {Umana}}]{2012ApJ...754..114H}
{Hayashida}, M., {Madejski}, G.~M., {Nalewajko}, K., {et~al.} 2012, \apj, 754,
  114

\bibitem[{{Healey} {et~al.}(2007){Healey}, {Romani}, {Taylor}, {Sadler},
  {Ricci}, {Murphy}, {Ulvestad}, \& {Winn}}]{2007ApJS..171...61H}
{Healey}, S.~E., {Romani}, R.~W., {Taylor}, G.~B., {et~al.} 2007, \apjs, 171,
  61

\bibitem[{{Hovatta} {et~al.}(2007){Hovatta}, {Tornikoski}, {Lainela}, {Lehto},
  {Valtaoja}, {Torniainen}, {Aller}, \& {Aller}}]{2007A&A...469..899H}
{Hovatta}, T., {Tornikoski}, M., {Lainela}, M., {et~al.} 2007, \aap, 469, 899

\bibitem[{{Hughes} {et~al.}(1992){Hughes}, {Aller}, \&
  {Aller}}]{1992ApJ...396..469H}
{Hughes}, P.~A., {Aller}, H.~D., \& {Aller}, M.~F. 1992, \apj, 396, 469

\bibitem[{{Kankkunen} {et~al.}(2025){Kankkunen}, {Tornikoski}, {Hovatta}, \&
  {L{\"a}hteenm{\"a}ki}}]{2025A&A...693A.318K}
{Kankkunen}, S., {Tornikoski}, M., {Hovatta}, T., \& {L{\"a}hteenm{\"a}ki}, A.
  2025, \aap, 693, A318

\bibitem[{{Larionov} {et~al.}(2008){Larionov}, {Jorstad}, {Marscher},
  {Raiteri}, {Villata}, {Agudo}, {Aller}, {Arkharov}, {Asfandiyarov}, {Bach},
  {Bachev}, {Berdyugin}, {B{\"o}ttcher}, {Buemi}, {Calcidese}, {Carosati},
  {Charlot}, {Chen}, {di Paola}, {Dolci}, {Dogru}, {Doroshenko}, {Efimov},
  {Erdem}, {Frasca}, {Fuhrmann}, {Giommi}, {Glowienka}, {Gupta}, {Gurwell},
  {Hagen-Thorn}, {Hsiao}, {Ibrahimov}, {Jordan}, {Kamada}, {Konstantinova},
  {Kopatskaya}, {Kovalev}, {Kovalev}, {Kurtanidze}, {L{\"a}hteenm{\"a}ki},
  {Lanteri}, {Larionova}, {Leto}, {Le Campion}, {Lee}, {Lindfors}, {Marilli},
  {McHardy}, {Mingaliev}, {Nazarov}, {Nieppola}, {Nilsson}, {Ohlert},
  {Pasanen}, {Porter}, {Pursimo}, {Ros}, {Sadakane}, {Sadun}, {Sergeev},
  {Smith}, {Strigachev}, {Sumitomo}, {Takalo}, {Tanaka}, {Trigilio}, {Umana},
  {Ungerechts}, {Volvach}, \& {Yuan}}]{2008A&A...492..389L}
{Larionov}, V.~M., {Jorstad}, S.~G., {Marscher}, A.~P., {et~al.} 2008, \aap,
  492, 389

\bibitem[{{Leto} {et~al.}(2009){Leto}, {Umana}, {Trigilio}, {Buemi}, {Dolei},
  {Manzitto}, {Cerrigone}, \& {Siringo}}]{2009A&A...507.1467L}
{Leto}, P., {Umana}, G., {Trigilio}, C., {et~al.} 2009, \aap, 507, 1467

\bibitem[{{Liodakis} {et~al.}(2017{\natexlab{a}}){Liodakis}, {Marchili},
  {Angelakis}, {Fuhrmann}, {Nestoras}, {Myserlis}, {Karamanavis}, {Krichbaum},
  {Sievers}, {Ungerechts}, \& {Zensus}}]{2017MNRAS.466.4625L}
{Liodakis}, I., {Marchili}, N., {Angelakis}, E., {et~al.} 2017{\natexlab{a}},
  \mnras, 466, 4625

\bibitem[{{Liodakis} {et~al.}(2017{\natexlab{b}}){Liodakis}, {Pavlidou},
  {Hovatta}, {Max-Moerbeck}, {Pearson}, {Richards}, \&
  {Readhead}}]{2017MNRAS.467.4565L}
{Liodakis}, I., {Pavlidou}, V., {Hovatta}, T., {et~al.} 2017{\natexlab{b}},
  \mnras, 467, 4565

\bibitem[{{MAGIC Collaboration} {et~al.}(2024){MAGIC Collaboration}, {Abe},
  {Abe}, {Abhir}, {Acciari}, {Agudo}, {Aniello}, {Ansoldi}, {Antonelli}, {Arbet
  Engels}, {Arcaro}, {Artero}, {Asano}, {Baack}, {Babi{\'c}}, {Baquero},
  {Barres de Almeida}, {Batkovi{\'c}}, {Baxter}, {Becerra Gonz{\'a}lez},
  {Bernardini}, {Bernete}, {Berti}, {Besenrieder}, {Bigongiari}, {Biland},
  {Blanch}, {Bonnoli}, {Bo{\v{s}}njak}, {Burelli}, {Busetto}, {Campoy-Ordaz},
  {Carosi}, {Carosi}, {Carretero-Castrillo}, {Castro-Tirado}, {Chai},
  {Cifuentes}, {Cikota}, {Colombo}, {Contreras}, {Cortina}, {Covino},
  {D'Amico}, {D'Elia}, {da Vela}, {Dazzi}, {de Angelis}, {de Lotto}, {Del
  Popolo}, {Delfino}, {Delgado}, {Delgado Mendez}, {Depaoli}, {di Pierro}, {di
  Venere}, {Dominis Prester}, {Donini}, {Dorner}, {Doro}, {Elsaesser}, {Emery},
  {Escudero}, {Fari{\~n}a}, {Fattorini}, {Foffano}, {Font}, {Fukami},
  {Fukazawa}, {Garc{\'\i}a L{\'o}pez}, {Gasparyan}, {Gaug}, {Giesbrecht Paiva},
  {Giglietto}, {Giordano}, {Gliwny}, {Grau}, {Green}, {Hadasch}, {Hahn},
  {Heckmann}, {Herrera}, {Hovatta}, {Hrupec}, {H{\"u}tten}, {Imazawa}, {Inada},
  {Iotov}, {Ishio}, {Jimenez Mart{\'\i}nez}, {Jormanainen}, {Kerszberg},
  {Kluge}, {Kobayashi}, {Kouch}, {Kubo}, {Kushida}, {L{\'a}inez Lez{\'a}un},
  {Lamastra}, {Leone}, {Lindfors}, {Liodakis}, {Lombardi}, {Longo},
  {L{\'o}pez-Moya}, {L{\'o}pez-Oramas}, {Loporchio}, {Lorini}, {Machado de
  Oliveira Fraga}, {Majumdar}, {Makariev}, {Maneva}, {Mang}, {Manganaro},
  {Mannheim}, {Mariotti}, {Mart{\'\i}nez}, {Mart{\'\i}nez-Chicharro},
  {Mas-Aguilar}, {Mazin}, {Menchiari}, {Mender}, {Miceli}, {Miener}, {Miranda},
  {Mirzoyan}, {Molero Gonz{\'a}lez}, {Molina}, {Mondal}, {Moralejo},
  {Morcuende}, {Nakamori}, {Nanci}, {Neustroev}, {Nigro}, {Nikoli{\'c}},
  {Nilsson}, {Nishijima}, {Njoh Ekoume}, {Noda}, {Nozaki}, {Ohtani}, {Okumura},
  {Otero-Santos}, {Paiano}, {Palatiello}, {Paneque}, {Paoletti}, {Paredes},
  {Pavlovi{\'c}}, {Persic}, {Pihet}, {Pirola}, {Podobnik}, {Prada Moroni},
  {Prandini}, {Principe}, {Priyadarshi}, {Rhode}, {Rib{\'o}}, {Rico}, {Righi},
  {Sahakyan}, {Saito}, {Satalecka}, {Saturni}, {Schleicher}, {Schmidt},
  {Schmuckermaier}, {Schubert}, {Schweizer}, {Sciaccaluga}, {Sitarek},
  {Spolon}, {Stamerra}, {Stri{\v{s}}kovi{\'c}}, {Strom}, {Suda}, {Suutarinen},
  {Tajima}, {Takeishi}, {Tavecchio}, {Temnikov}, {Terauchi}, {Terzi{\'c}},
  {Teshima}, {Tosti}, {Truzzi}, {Tutone}, {Ubach}, {van Scherpenberg},
  {Ventura}, {Verguilov}, {Viale}, {Vigorito}, {Vitale}, {Walter},
  {Wunderlich}, {Yamamoto}, \& {MWL Collaborators}}]{2024MNRAS.529.3894M}
{MAGIC Collaboration}, {Abe}, H., {Abe}, S., {et~al.} 2024, \mnras, 529, 3894

\bibitem[{{Massaro} {et~al.}(2015){Massaro}, {Maselli}, {Leto}, {Marchegiani},
  {Perri}, {Giommi}, \& {Piranomonte}}]{2015Ap&SS.357...75M}
{Massaro}, E., {Maselli}, A., {Leto}, C., {et~al.} 2015, \apss, 357, 75

\bibitem[{{Otero-Santos} {et~al.}(2025){Otero-Santos}, {Raiteri}, {Tramacere},
  {Escudero Pedrosa}, {Acosta-Pulido}, {Carnerero}, {Villata}, {Agudo},
  {Rahimov}, {Andreeva}, {Ivanov}, {Marchili}, {Righini}, {Giroletti},
  {Gurwell}, {Savchenko}, {Carosati}, {Chen}, {Kurtanidze}, {Joner}, {Semkov},
  {Pursimo}, {Ben{\'\i}tez}, {Damljanovic}, {Andreuzzi}, {Apolonio}, {Borman},
  {Bozhilov}, {Galindo-Guil}, {Grishina}, {Hagen-Thorn}, {Hiriart}, {Hsiao},
  {Ibryamov}, {Ivanidze}, {Kimeridze}, {Kopatskaya}, {Kurtanidze}, {Larionov},
  {Larionova}, {Larionova}, {Minev}, {Morozova}, {Nikolashvili}, {Ovcharov},
  {Sigua}, {Stojanovic}, {Troitskiy}, {Troitskaya}, {Tsai}, {Valcheva},
  {Vasilyev}, {Vince}, {Zaharieva}, \& {Zhovtan}}]{2025A&A...693A.196O}
{Otero-Santos}, J., {Raiteri}, C.~M., {Tramacere}, A., {et~al.} 2025, \aap,
  693, A196

\bibitem[{{Perley} \& {Butler}(2013)}]{2013ApJS..204...19P}
{Perley}, R.~A. \& {Butler}, B.~J. 2013, \apjs, 204, 19

\bibitem[{{Raiteri} {et~al.}(2017){Raiteri}, {Villata}, {Acosta-Pulido},
  {Agudo}, {Arkharov}, {Bachev}, {Baida}, {Ben{\'\i}tez}, {Borman}, {Boschin},
  {Bozhilov}, {Butuzova}, {Calcidese}, {Carnerero}, {Carosati}, {Casadio},
  {Castro-Segura}, {Chen}, {Damljanovic}, {D'Ammando}, {di Paola},
  {Echevarr{\'\i}a}, {Efimova}, {Ehgamberdiev}, {Espinosa}, {Fuentes},
  {Giunta}, {G{\'o}mez}, {Grishina}, {Gurwell}, {Hiriart}, {Jermak}, {Jordan},
  {Jorstad}, {Joshi}, {Kopatskaya}, {Kuratov}, {Kurtanidze}, {Kurtanidze},
  {L{\"a}hteenm{\"a}ki}, {Larionov}, {Larionova}, {Larionova}, {L{\'a}zaro},
  {Lin}, {Malmrose}, {Marscher}, {Matsumoto}, {McBreen}, {Michel}, {Mihov},
  {Minev}, {Mirzaqulov}, {Mokrushina}, {Molina}, {Moody}, {Morozova},
  {Nazarov}, {Nikolashvili}, {Ohlert}, {Okhmat}, {Ovcharov}, {Pinna},
  {Polakis}, {Protasio}, {Pursimo}, {Redondo-Lorenzo}, {Rizzi},
  {Rodriguez-Coira}, {Sadakane}, {Sadun}, {Samal}, {Savchenko}, {Semkov},
  {Skiff}, {Slavcheva-Mihova}, {Smith}, {Steele}, {Strigachev}, {Tammi},
  {Thum}, {Tornikoski}, {Troitskaya}, {Troitsky}, {Vasilyev}, \&
  {Vince}}]{2017Natur.552..374R}
{Raiteri}, C.~M., {Villata}, M., {Acosta-Pulido}, J.~A., {et~al.} 2017, \nat,
  552, 374

\bibitem[{{Raiteri} {et~al.}(2009){Raiteri}, {Villata}, {Capetti}, {Aller},
  {Bach}, {Calcidese}, {Gurwell}, {Larionov}, {Ohlert}, {Nilsson},
  {Strigachev}, {Agudo}, {Aller}, {Bachev}, {Ben{\'\i}tez}, {Berdyugin},
  {B{\"o}ttcher}, {Buemi}, {Buttiglione}, {Carosati}, {Charlot}, {Chen},
  {Dultzin}, {Forn{\'e}}, {Fuhrmann}, {G{\'o}mez}, {Gupta}, {Heidt}, {Hiriart},
  {Hsiao}, {Jel{\'\i}nek}, {Jorstad}, {Kimeridze}, {Konstantinova},
  {Kopatskaya}, {Kostov}, {Kurtanidze}, {L{\"a}hteenm{\"a}ki}, {Lanteri},
  {Larionova}, {Leto}, {Latev}, {Le Campion}, {Lee}, {Ligustri}, {Lindfors},
  {Marscher}, {Mihov}, {Nikolashvili}, {Nikolov}, {Ovcharov}, {Principe},
  {Pursimo}, {Ragozzine}, {Robb}, {Ros}, {Sadun}, {Sagar}, {Semkov}, {Sigua},
  {Smart}, {Sorcia}, {Takalo}, {Tornikoski}, {Trigilio}, {Uckert}, {Umana},
  {Valcheva}, \& {Volvach}}]{2009A&A...507..769R}
{Raiteri}, C.~M., {Villata}, M., {Capetti}, A., {et~al.} 2009, \aap, 507, 769

\bibitem[{{Raiteri} {et~al.}(2024){Raiteri}, {Villata}, {Carnerero},
  {Kurtanidze}, {Mirzaqulov}, {Ben{\'\i}tez}, {Bonnoli}, {Carosati},
  {Acosta-Pulido}, {Agudo}, {Andreeva}, {Apolonio}, {Bachev}, {Borman},
  {Bozhilov}, {Brown}, {Carbonell}, {Casadio}, {Chen}, {Damljanovic},
  {Ehgamberdiev}, {Elsaesser}, {Escudero}, {Feige}, {Fuentes}, {Gabellini},
  {Gazeas}, {Giroletti}, {Grishina}, {Gupta}, {Gurwell}, {Hagen-Thorn},
  {Hamed}, {Hiriart}, {Hodges}, {Ivanidze}, {Ivanov}, {Joner}, {Jorstad},
  {Jovanovic}, {Kiehlmann}, {Kimeridze}, {Kopatskaya}, {Kovalev}, {Kovalev},
  {Kurtanidze}, {Kurtenkov}, {Larionova}, {Lessing}, {Lin}, {L{\'o}pez},
  {Lorey}, {Ludwig}, {Marchili}, {Marchini}, {Marscher}, {Matsumoto},
  {Max-Moerbeck}, {Mihov}, {Minev}, {Mingaliev}, {Modaressi}, {Morozova},
  {Mortari}, {Mufakharov}, {Myserlis}, {Nikolashvili}, {Pearson}, {Popkov},
  {Rahimov}, {Readhead}, {Reinhart}, {Reeves}, {Righini}, {Romanov},
  {Savchenko}, {Semkov}, {Shishkina}, {Sigua}, {Slavcheva-Mihova}, {Sotnikova},
  {Steineke}, {Stojanovic}, {Strigachev}, {Takey}, {Traianou}, {Troitskaya},
  {Troitskiy}, {Tsai}, {Valcheva}, {Vasilyev}, {Verna}, {Vince}, {Vrontaki},
  {Weaver}, {Webb}, {Yuldoshev}, {Zaharieva}, \&
  {Zhovtan}}]{2024A&A...692A..48R}
{Raiteri}, C.~M., {Villata}, M., {Carnerero}, M.~I., {et~al.} 2024, \aap, 692,
  A48

\bibitem[{{Raiteri} {et~al.}(2021){Raiteri}, {Villata}, {Larionov}, {Jorstad},
  {Marscher}, {Weaver}, {Acosta-Pulido}, {Agudo}, {Andreeva}, {Arkharov},
  {Bachev}, {Ben{\'\i}tez}, {Berton}, {Bj{\"o}rklund}, {Borman}, {Bozhilov},
  {Carnerero}, {Carosati}, {Casadio}, {Chen}, {Damljanovic}, {D'Ammando},
  {Escudero}, {Fuentes}, {Giroletti}, {Grishina}, {Gupta}, {Hagen-Thorn},
  {Hart}, {Hiriart}, {Hou}, {Ivanov}, {Kim}, {Kimeridze}, {Konstantopoulou},
  {Kopatskaya}, {Kurtanidze}, {Kurtanidze}, {L{\"a}hteenm{\"a}ki}, {Larionova},
  {Larionova}, {Marchili}, {Markovic}, {Minev}, {Morozova}, {Myserlis},
  {Nakamura}, {Nikiforova}, {Nikolashvili}, {Otero-Santos}, {Ovcharov},
  {Pursimo}, {Rahimov}, {Righini}, {Sakamoto}, {Savchenko}, {Semkov},
  {Shakhovskoy}, {Sigua}, {Stojanovic}, {Strigachev}, {Thum}, {Tornikoski},
  {Traianou}, {Troitskaya}, {Troitskiy}, {Tsai}, {Valcheva}, {Vasilyev},
  {Vince}, \& {Zaharieva}}]{2021MNRAS.504.5629R}
{Raiteri}, C.~M., {Villata}, M., {Larionov}, V.~M., {et~al.} 2021, \mnras, 504,
  5629

\bibitem[{{Raiteri} {et~al.}(2007){Raiteri}, {Villata}, {Larionov}, {Pursimo},
  {Ibrahimov}, {Nilsson}, {Aller}, {Kurtanidze}, {Foschini}, {Ohlert},
  {Papadakis}, {Sumitomo}, {Volvach}, {Aller}, {Arkharov}, {Bach}, {Berdyugin},
  {B{\"o}ttcher}, {Buemi}, {Calcidese}, {Charlot}, {Delgado S{\'a}nchez}, {di
  Paola}, {Djupvik}, {Dolci}, {Efimova}, {Fan}, {Forn{\'e}}, {Gomez}, {Gupta},
  {Hagen-Thorn}, {Hooks}, {Hovatta}, {Ishii}, {Kamada}, {Konstantinova},
  {Kopatskaya}, {Kovalev}, {Kovalev}, {L{\"a}hteenm{\"a}ki}, {Lanteri}, {Le
  Campion}, {Lee}, {Leto}, {Lin}, {Lindfors}, {Mingaliev}, {Mizoguchi},
  {Nicastro}, {Nikolashvili}, {Nishiyama}, {{\"O}stman}, {Ovcharov},
  {P{\"a}{\"a}kk{\"o}nen}, {Pasanen}, {Pian}, {Rector}, {Ros}, {Sadakane},
  {Selj}, {Semkov}, {Sharapov}, {Somero}, {Stanev}, {Strigachev}, {Takalo},
  {Tanaka}, {Tavani}, {Torniainen}, {Tornikoski}, {Trigilio}, {Umana},
  {Vercellone}, {Valcheva}, {Volvach}, \& {Yamanaka}}]{2007A&A...473..819R}
{Raiteri}, C.~M., {Villata}, M., {Larionov}, V.~M., {et~al.} 2007, \aap, 473,
  819

\bibitem[{{Richards} {et~al.}(2011){Richards}, {Max-Moerbeck}, {Pavlidou},
  {King}, {Pearson}, {Readhead}, {Reeves}, {Shepherd}, {Stevenson},
  {Weintraub}, {Fuhrmann}, {Angelakis}, {Zensus}, {Healey}, {Romani}, {Shaw},
  {Grainge}, {Birkinshaw}, {Lancaster}, {Worrall}, {Taylor}, {Cotter}, \&
  {Bustos}}]{2011ApJS..194...29R}
{Richards}, J.~L., {Max-Moerbeck}, W., {Pavlidou}, V., {et~al.} 2011, \apjs,
  194, 29

\bibitem[{{Simonetti} {et~al.}(1985){Simonetti}, {Cordes}, \&
  {Heeschen}}]{1985ApJ...296...46S}
{Simonetti}, J.~H., {Cordes}, J.~M., \& {Heeschen}, D.~S. 1985, \apj, 296, 46

\bibitem[{{Venturi} {et~al.}(2001){Venturi}, {Dallacasa}, {Orfei}, {Bondi},
  {Fanti}, {Gregorini}, {Mantovani}, {Stanghellini}, {Trigilio}, \&
  {Umana}}]{2001A&A...379..755V}
{Venturi}, T., {Dallacasa}, D., {Orfei}, A., {et~al.} 2001, \aap, 379, 755

\bibitem[{{Vercellone} {et~al.}(2010){Vercellone}, {D'Ammando}, {Vittorini},
  {Donnarumma}, {Pucella}, {Tavani}, {Ferrari}, {Raiteri}, {Villata}, {Romano},
  {Krimm}, {Tiengo}, {Chen}, {Giovannini}, {Venturi}, {Giroletti}, {Kovalev},
  {Sokolovsky}, {Pushkarev}, {Lister}, {Argan}, {Barbiellini}, {Bulgarelli},
  {Caraveo}, {Cattaneo}, {Cocco}, {Costa}, {Del Monte}, {De Paris}, {Di Cocco},
  {Evangelista}, {Feroci}, {Fiorini}, {Fornari}, {Froysland}, {Fuschino},
  {Galli}, {Gianotti}, {Labanti}, {Lapshov}, {Lazzarotto}, {Lipari}, {Longo},
  {Giuliani}, {Marisaldi}, {Mereghetti}, {Morselli}, {Pellizzoni}, {Pacciani},
  {Perotti}, {Piano}, {Picozza}, {Pilia}, {Prest}, {Rapisarda}, {Rappoldi},
  {Sabatini}, {Soffitta}, {Striani}, {Trifoglio}, {Trois}, {Vallazza},
  {Zambra}, {Zanello}, {Pittori}, {Verrecchia}, {Santolamazza}, {Giommi},
  {Colafrancesco}, {Salotti}, {Agudo}, {Aller}, {Aller}, {Arkharov}, {Bach},
  {Bachev}, {Beltrame}, {Ben{\'\i}tez}, {B{\"o}ttcher}, {Buemi}, {Calcidese},
  {Capezzali}, {Carosati}, {Chen}, {Da Rio}, {Di Paola}, {Dolci}, {Dultzin},
  {Forn{\'e}}, {G{\'o}mez}, {Gurwell}, {Hagen-Thorn}, {Halkola}, {Heidt},
  {Hiriart}, {Hovatta}, {Hsiao}, {Jorstad}, {Kimeridze}, {Konstantinova},
  {Kopatskaya}, {Koptelova}, {Kurtanidze}, {L{\"a}hteenm{\"a}ki}, {Larionov},
  {Leto}, {Ligustri}, {Lindfors}, {Lopez}, {Marscher}, {Mujica},
  {Nikolashvili}, {Nilsson}, {Mommert}, {Palma}, {Pasanen}, {Roca-Sogorb},
  {Ros}, {Roustazadeh}, {Sadun}, {Saino}, {Sigua}, {Sorcia}, {Takalo},
  {Tornikoski}, {Trigilio}, {Turchetti}, \& {Umana}}]{2010ApJ...712..405V}
{Vercellone}, S., {D'Ammando}, F., {Vittorini}, V., {et~al.} 2010, \apj, 712,
  405

\bibitem[{{Villata} {et~al.}(2006){Villata}, {Raiteri}, {Balonek}, {Aller},
  {Jorstad}, {Kurtanidze}, {Nicastro}, {Nilsson}, {Aller}, {Arai}, {Arkharov},
  {Bach}, {Ben{\'{\i}}tez}, {Berdyugin}, {Buemi}, {B{\"o}ttcher}, {Carosati},
  {Casas}, {Caulet}, {Chen}, {Chiang}, {Chou}, {Ciprini}, {Coloma}, {di Rico},
  {D{\'{\i}}az}, {Efimova}, {Forsyth}, {Frasca}, {Fuhrmann}, {Gadway}, {Gupta},
  {Hagen-Thorn}, {Harvey}, {Heidt}, {Hernandez-Toledo}, {Hroch}, {Hu}, {Hudec},
  {Ibrahimov}, {Imada}, {Kamata}, {Kato}, {Katsuura}, {Konstantinova},
  {Kopatskaya}, {Kotaka}, {Kovalev}, {Kovalev}, {Krichbaum}, {Kubota},
  {Kurosaki}, {Lanteri}, {Larionov}, {Larionova}, {Laurikainen}, {Lee}, {Leto},
  {L{\"a}hteenm{\"a}ki}, {L{\'o}pez-Cruz}, {Marilli}, {Marscher}, {McHardy},
  {Mondal}, {Mullan}, {Napoleone}, {Nikolashvili}, {Ohlert}, {Postnikov},
  {Pursimo}, {Ragni}, {Ros}, {Sadakane}, {Sadun}, {Savolainen}, {Sergeeva},
  {Sigua}, {Sillanp{\"a}{\"a}}, {Sixtova}, {Sumitomo}, {Takalo},
  {Ter{\"a}sranta}, {Tornikoski}, {Trigilio}, {Umana}, {Volvach}, {Voss}, \&
  {Wortel}}]{2006A&A...453..817V}
{Villata}, M., {Raiteri}, C.~M., {Balonek}, T.~J., {et~al.} 2006, \aap, 453,
  817

\bibitem[{{Villata} {et~al.}(2002){Villata}, {Raiteri}, {Kurtanidze},
  {Nikolashvili}, {Ibrahimov}, {Papadakis}, {Tsinganos}, {Sadakane}, {Okada},
  {Takalo}, {Sillanp{\"a}{\"a}}, {Tosti}, {Ciprini}, {Frasca}, {Marilli},
  {Robb}, {Noble}, {Jorstad}, {Hagen-Thorn}, {Larionov}, {Nesci}, {Maesano},
  {Schwartz}, {Basler}, {Gorham}, {Iwamatsu}, {Kato}, {Pullen}, {Ben{\'\i}tez},
  {de Diego}, {Moilanen}, {Oksanen}, {Rodriguez}, {Sadun}, {Kelly}, {Carini},
  {Miller}, {Catalano}, {Dultzin-Hacyan}, {Fan}, {Ishioka}, {Karttunen},
  {Kein{\"a}nen}, {Kudryavtseva}, {Lainela}, {Lanteri}, {Larionova},
  {Matsumoto}, {Mattox}, {Montagni}, {Nucciarelli}, {Ostorero},
  {Papamastorakis}, {Pasanen}, {Sobrito}, \& {Uemura}}]{2002A&A...390..407V}
{Villata}, M., {Raiteri}, C.~M., {Kurtanidze}, O.~M., {et~al.} 2002, \aap, 390,
  407

\end{thebibliography}

\onecolumn

\begin{appendix}

\section{Tables}
\label{sec:app1}

\begin{table}[hbt!]
\caption{ Basic information about the sources included in the Blazar IRA Monitoring Program.}
\label{table:1}
 \FloatBarrier
\centering  
\scalebox{0.97}{
\begin{tabular}{lcccccrcrcrcrc} 
\hline\hline 
Source & Alternative & Blazar & \multicolumn{2}{c}{Coordinates (J2000)} & $z$ & \multicolumn{2}{c}{5 GHz}  & \multicolumn{2}{c}{8 GHz} & \multicolumn{2}{c}{24 GHz} & \multicolumn{2}{c}{43 GHz}\\
Name & Name & Type & RA & Dec & & $N$  & $\langle S\rangle$ & $N$  & $\langle S\rangle$ & $N$  & $\langle S\rangle$ & $N$  & $\langle S\rangle$ \\
 & & & & & & & (Jy) & & (Jy) & & (Jy) & & (Jy)\\
(1) & (2) & (3) & (4) & (5) & (6) & (7) & (8) & (9) & (10) & (11) & (12) & (13) & (14) \\
\hline
0219+428  &  3C 66A  &  BL Lac  &  02:22:39.6 & +43:02:07.8 & 0.37 &  36 &    1.82 & 121 &    1.06 &  81 &    0.95 &  36 &    1.05\\
0235+164  &  AO 0235+16  &  BL Lac  &  02:38:38.9 & +16:36:59.3 & 0.94 &  59 &    1.54 & 142 &    1.67 & 114 &    1.96 &  45 &    2.32\\
0316+413  &  3C 84  &  uncertain  &  03:19:48.2 & +41:30:42.1 & 0.02 &   4 &   30.56 &  54 &   36.20 &  50 &   26.78 &   0 &   ---\\
0323+342  &  B2 0321+33B  &  uncertain  &  03:24:41.2 & +34:10:45.9 & 0.06 &   0 &   --- &  18 &    0.34 &  12 &    0.37 &   0 &   ---\\
0336--019  &  CTA 026  &  FSRQ  &  03:39:30.9 & --01:46:35.8 & 0.85 &  57 &    2.69 & 136 &    2.02 & 105 &    1.87 &  39 &    1.89\\
0355+508  &  NRAO 150  &  FSRQ  &  03:59:29.7 & +50:57:50.2 & 1.52 &  55 &    7.59 & 124 &    9.03 &  88 &    9.59 &  32 &   11.31\\
0420--014  &  PKS 0420--01  &  FSRQ  &  04:23:15.8 & --01:20:33.1 & 0.92 &  59 &    3.54 & 138 &    3.73 & 114 &    4.54 &  42 &    4.44\\
0440--003  &  NRAO 190  &  FSRQ  &  04:42:38.6 & --00:17:43.4 & 0.84 &  45 &    2.44 & 114 &    1.70 &  94 &    1.09 &  35 &    0.92\\
0506+056  &  TXS 0506+056  &  BL Lac  &  05:09:26.0 & +05:41:35.3 & 0.34 &   4 &    1.44 &  49 &    1.36 &  48 &    1.35 &   0 &   ---\\
0528+134  &  PKS 0528+134  &  FSRQ  &  05:30:56.4 & +13:31:55.1 & 2.06 &  61 &    2.22 & 133 &    1.99 & 109 &    2.05 &  43 &    3.22\\
0716+714  &  S5 0716+71  &  BL Lac  &  07:21:53.4 & +71:20:36.4 & 0.31 &  70 &    1.16 & 144 &    1.24 & 109 &    1.66 &  43 &    2.73\\
0735+178  &  PKS 0735+17  &  BL Lac  &  07:38:07.3 & +17:42:19.0 & 0.42 &  55 &    1.09 & 138 &    0.97 &  91 &    0.78 &  37 &    0.69\\
0736+017  &  PKS 0736+01  &  FSRQ  &  07:39:18.0 & +01:37:04.6 & 0.19 &  45 &    1.17 & 132 &    1.32 &  99 &    1.69 &  38 &    1.43\\
0827+243  &  OJ 248  &  FSRQ  &  08:30:52.1 & +24:10:59.8 & 0.94 &  38 &    0.96 & 122 &    0.93 &  78 &    1.02 &  22 &    1.38\\
0829+046  &  OJ 49  &  BL Lac  &  08:31:48.9 & +04:29:39.1 & 0.17 &  55 &    0.67 & 141 &    0.90 &  90 &    1.22 &  35 &    0.56\\
0836+710  &  S5 0836+71  &  FSRQ  &  08:41:24.3 & +70:53:42.2 & 2.17 &  67 &    2.59 & 143 &    2.24 & 104 &    2.02 &  37 &    1.94\\
0838+133  &  3C 207  &  uncertain  &  08:40:47.6 & +13:12:23.6 & 0.68 &  16 &    1.70 &  84 &    1.54 &  41 &    1.37 &   0 &   ---\\
0851+202  &  OJ 287  &  BL Lac  &  08:54:48.8 & +20:06:30.6 & 0.31 &  69 &    3.01 & 145 &    4.23 &  95 &    4.93 &  44 &    4.40\\
0954+658  &  S4 0954+65  &  BL Lac  &  09:58:47.2 & +65:33:54.8 & 0.37 &  60 &    1.01 & 147 &    1.10 & 105 &    1.43 &  40 &    1.55\\
1101+384  &  MRK 421  &  BL Lac  &  11:04:27.3 & +38:12:31.8 & 0.03 &  47 &    0.68 &  99 &    0.57 &  43 &    0.44 &   0 &   ---\\
1156+295  &  4C +29.45  &  FSRQ  &  11:59:31.8 & +29:14:43.8 & 0.72 &  65 &    1.66 & 136 &    2.05 &  99 &    2.62 &  33 &    2.06\\
1219+285  &  W Com  &  BL Lac  &  12:21:31.7 & +28:13:58.5 & 0.10 &  42 &    0.55 & 126 &    0.55 &  50 &    0.49 &  33 &    0.48\\
1222+21  &  PKS 1222+216  &  FSRQ  &  12:24:54.4 & +21:22:46.4 & 0.43 &   2 &    1.95 &  43 &    1.74 &  30 &    1.36 &   0 &   ---\\
1226+023  &  3C 273  &  FSRQ  &  12:29:06.7 & +02:03:08.6 & 0.16 &  74 &   33.86 & 150 &   26.80 & 118 &   21.28 &  44 &   21.37\\
1253--055  &  3C 279  &  FSRQ  &  12:56:11.1 & --05:47:21.5 & 0.54 &  73 &   13.89 & 146 &   16.21 & 113 &   18.01 &  44 &   18.08\\
1334--127  &  PKS 1335--127  &  FSRQ  &  13:37:39.8 & --12:57:24.7 & 0.54 &  58 &    3.91 & 127 &    4.20 & 103 &    4.67 &  40 &    6.35\\
1354+195  &  4C +19.44  &  FSRQ  &  13:57:04.4 & +19:19:07.4 & 0.72 &  46 &    1.86 & 131 &    1.89 &  95 &    1.87 &  41 &    1.93\\
1441+252  &  PKS 1441+25  &  FSRQ  &  14:43:56.9 & +25:01:44.5 & 0.94 &   1 &    0.24 &  42 &    0.25 &  34 &    0.18 &   0 &   ---\\
1510--089  &  PKS 1510--08  &  FSRQ  &  15:12:50.5 & --09:05:59.8 & 0.36 &  68 &    2.69 & 146 &    2.82 & 103 &    3.08 &  35 &    2.17\\
1553+113  &  PG 1553+113  &  BL Lac  &  15:55:43.0 & +11:11:24.4 & 0.36 &   9 &    0.26 &  66 &    0.28 &  53 &    0.30 &   0 &   ---\\
1606+105  &  4C +10.45  &  FSRQ  &  16:08:46.1 & +10:29:07.8 & 1.23 &  56 &    1.23 & 133 &    1.22 &  84 &    1.04 &  31 &    1.15\\
1611+343  &  B2 1611+34  &  FSRQ  &  16:13:41.0 & +34:12:47.9 & 1.40 &  66 &    4.12 & 144 &    4.15 & 106 &    3.26 &  38 &    1.99\\
1633+382  &  S4 1633+38  &  FSRQ  &  16:35:15.4 & +38:08:04.5 & 1.81 &  63 &    2.81 & 147 &    2.93 & 104 &    3.03 &  36 &    2.92\\
1641+399  &  3C 345  &  FSRQ  &  16:42:58.8 & +39:48:37.0 & 0.59 &  65 &    6.59 & 150 &    6.02 & 107 &    6.06 &  34 &    5.20\\
1652+398  &  MRK 501  &  BL Lac  &  16:53:52.2 & +39:45:36.6 & 0.03 &  37 &    1.45 & 104 &    1.29 &  54 &    1.03 &   0 &   ---\\
1730--130  &  NRAO 530  &  FSRQ  &  17:33:02.6 & --13:04:49.5 & 0.90 &  51 &    4.47 & 124 &    4.41 &  95 &    3.68 &  36 &    3.79\\
1741--038  &  PKS 1741--03  &  FSRQ  &  17:43:58.8 & --03:50:04.6 & 1.05 &  56 &    4.28 & 133 &    4.64 & 103 &    4.30 &  37 &    2.73\\
1807+698  &  3C 371  &  BL Lac  &  18:06:50.7 & +69:49:28.1 & 0.05 &  50 &    1.78 & 122 &    1.84 &  93 &    1.80 &  37 &    1.41\\
1828+487  &  3C 380  &  uncertain  &  18:29:31.8 & +48:44:46.1 & 0.69 &  23 &    4.98 &  86 &    3.81 &  55 &    2.33 &   0 &   ---\\
1830--211  &  PKS 1830--21  &  FSRQ  &  18:33:39.9 & --21:03:39.8 & 2.51 &   0 &   --- &  29 &    9.73 &  25 &    6.01 &   0 &   ---\\
1959+650  &  1ES 1959+650  &  BL Lac  &  19:59:59.8 & +65:08:54.6 & 0.05 &   2 &    0.27 &  44 &    0.25 &  36 &    0.24 &   0 &   ---\\
2200+420  &  BL Lac  &  BL Lac  &  22:02:43.2 & +42:16:40.0 & 0.07 &  74 &    3.95 & 151 &    3.87 & 115 &    3.89 &  40 &    3.40\\
2223--052  &  3C 446  &  FSRQ  &  22:25:47.2 & --04:57:01.4 & 1.40 &  52 &    5.31 & 131 &    4.85 &  98 &    4.49 &  38 &    6.97\\
2230+114  &  CTA 102  &  FSRQ  &  22:32:36.4 & +11:43:50.9 & 1.03 &  67 &    4.42 & 144 &    3.87 & 104 &    3.98 &  39 &    4.85\\
2251+158  &  3C 454.3  &  FSRQ  &  22:53:57.7 & +16:08:53.6 & 0.86 &  77 &   12.46 & 151 &   12.67 & 119 &   11.96 &  47 &   14.44\\
2344+092  &  4C 09.74  &  FSRQ  &  23:46:36.8 & +09:30:45.5 & 0.67 &  32 &    1.25 & 126 &    1.20 &  85 &    1.14 &  36 &    0.77\\
2344+514  &  1ES 2344+514  &  BL Lac  &  23:47:04.8 & +51:42:17.9 & 0.04 &   2 &    0.24 &  31 &    0.23 &  23 &    0.17 &   0 &   ---\\
\hline
\end{tabular}} 
\tablefoot{The Table reports:  the IAU name (Col. 1); an alternative (often more common) name (Col. 2); the blazar type (Col. 3); the coordinates in RA-Dec (Cols. 4 and 5); the redshift (Col.6); for each of the four monitored radio bands, the number of epochs in which the source was observed (Cols. 7, 9, 11, and 13), and the average flux density (Cols. 8, 10, 12, 14) calculated over these epochs.}
\end{table}

\begin{table}[hbt!]
\caption{ Basic characteristics of the monitored sources at different bands.}
\label{table:2}
\centering
\begin{tabular}{lccccccccccccc}
\hline\hline
Source & \multicolumn{3}{c}{5 GHz}  & \multicolumn{3}{c}{8 GHz} & \multicolumn{3}{c}{24 GHz} & \multicolumn{3}{c}{43 GHz} & $\alpha$\\
Name & $\overline{m}$ & $SF_{1.5}^\prime$ & $\frac{SF_{1.5}^\prime}{SF_{3.0}^\prime}$ & $\overline{m}$ & $SF_{1.5}^\prime$ & $\frac{SF_{1.5}^\prime}{SF_{3.0}^\prime}$ & $\overline{m}$ & $SF_{1.5}^\prime$ & $\frac{SF_{1.5}^\prime}{SF_{3.0}^\prime}$ & $\overline{m}$ & $SF_{1.5}^\prime$ & $\frac{SF_{1.5}^\prime}{SF_{3.0}^\prime}$ & \\
 & (\%) & (\%) & & (\%) & (\%) & & (\%) & (\%) & & (\%) & (\%) & & \\
 (1) & (2) & (3) & (4) & (5) & (6) & (7) & (8) & (9) & (10) & (11) & (12) & (13) & (14) \\
\hline
0219+428  &    24.70 &   27.74 &    1.37 &    17.87 &   14.99 &    0.98 &    19.65 &   23.59 &    1.03 &    10.23 &   19.25 &    0.80 & --0.11\\
0235+164  &    38.29 &   28.82 &    0.55 &    50.55 &   48.63 &    0.80 &    49.83 &   49.02 &    0.87 &    56.22 &   58.69 &    4.83 & +0.08\\
0316+413  &    --- &   --- &   --- &    19.03 &   13.78 &    1.09 &    13.06 &   13.53 &    1.48 &    --- &   --- &   --- & --0.29\\
0323+342  &    --- &   --- &   --- &    13.68 &   11.46 &    0.70 &     9.97 &   20.47 &    1.55 &    --- &   --- &   --- & +0.08\\
0336--019  &    22.28 &   19.55 &    0.86 &    39.29 &   17.81 &    0.58 &    34.97 &   32.13 &    0.78 &    18.87 &   23.28 &    0.91 & --0.01\\
0355+508  &    17.57 &   14.85 &    1.32 &    22.51 &   19.08 &    0.78 &    23.80 &   20.35 &    1.24 &    13.36 &   21.46 &   --- & +0.08\\
0420--014  &    28.12 &   23.53 &    0.66 &    34.57 &   24.25 &    0.57 &    42.62 &   30.30 &    0.62 &    25.16 &   44.39 &    0.96 & +0.21\\
0440--003  &    19.16 &    8.01 &    0.41 &    15.69 &   10.72 &    0.72 &    18.85 &   27.20 &    1.21 &    12.00 &   18.67 &    0.86 & --0.42\\
0506+056  &    --- &   --- &   --- &    31.53 &   35.54 &    0.91 &    34.62 &   38.87 &    0.98 &    --- &   --- &   --- & --0.03\\
0528+134  &    42.92 &   13.64 &    0.51 &    55.33 &   16.26 &    0.60 &    62.42 &   39.50 &    0.93 &    40.67 &   51.35 &    1.30 & +0.00\\
0716+714  &    35.55 &   25.83 &    0.83 &    40.08 &   31.44 &    0.88 &    56.81 &   53.39 &    0.86 &    75.48 &   91.36 &    0.96 & +0.23\\
0735+178  &    16.25 &   11.80 &    0.57 &    22.19 &   19.56 &    0.73 &    26.66 &   26.69 &    0.70 &    16.98 &   28.29 &    0.68 & --0.17\\
0736+017  &    22.05 &   21.07 &    0.91 &    27.48 &   30.89 &    1.05 &    30.48 &   38.47 &    1.09 &    25.59 &   35.81 &    1.60 & +0.18\\
0827+243  &    22.00 &   19.78 &    0.72 &    25.79 &   28.45 &    0.88 &    25.78 &   32.66 &    1.69 &    26.83 &   32.75 &   --- & +0.06\\
0829+046  &     9.40 &   11.09 &    0.72 &    32.02 &   25.01 &    1.16 &    42.06 &   40.79 &    1.13 &    25.53 &   33.20 &    0.62 & +0.19\\
0836+710  &    18.95 &   14.17 &    0.88 &    20.16 &   13.19 &    0.77 &    19.34 &   24.38 &    1.31 &    21.16 &   29.41 &    0.96 & --0.07\\
0838+133  &     4.47 &    6.54 &    1.61 &    11.07 &   11.71 &    1.09 &    13.93 &   18.02 &    0.95 &    --- &   --- &   --- & --0.14\\
0851+202  &    40.45 &   24.29 &    0.77 &    37.66 &   25.23 &    1.03 &    43.17 &   35.54 &    0.94 &    66.30 &   87.91 &    2.38 & +0.22\\
0954+658  &    16.71 &   18.97 &    1.08 &    24.68 &   24.23 &    0.87 &    32.75 &   37.69 &    1.07 &    19.03 &   24.68 &    0.75 & +0.25\\
1101+384  &    11.78 &   13.92 &    1.02 &    11.76 &   14.94 &    0.93 &     8.66 &   17.31 &    0.73 &    --- &   --- &   --- & --0.18\\
1156+295  &    29.74 &   42.17 &    1.84 &    48.67 &   50.46 &    1.06 &    55.81 &   58.52 &    0.87 &    37.40 &   49.77 &    2.28 & +0.09\\
1219+285  &    20.31 &   16.43 &    0.77 &    12.75 &   15.43 &    1.00 &    16.66 &   27.58 &    1.34 &    16.02 &   26.67 &    0.76 & --0.16\\
1222+21  &    --- &   --- &   --- &    29.46 &   19.92 &    0.48 &    30.22 &   21.73 &    0.45 &    --- &   --- &   --- & --0.34\\
1226+023  &    10.76 &    7.70 &    0.71 &    17.55 &    7.93 &    0.59 &    27.02 &   14.37 &    0.70 &    16.16 &   20.81 &    0.79 & --0.26\\
1253--055  &    21.40 &   15.81 &    0.79 &    21.65 &   16.73 &    0.67 &    22.65 &   20.59 &    0.64 &    15.26 &   18.70 &    0.63 & +0.12\\
1334--127  &    26.96 &   24.53 &    0.65 &    28.89 &   24.58 &    0.82 &    30.78 &   23.72 &    0.80 &    34.38 &   55.10 &    1.19 & +0.05\\
1354+195  &    10.21 &    5.06 &    0.57 &    10.29 &   10.92 &    0.65 &    15.17 &   18.89 &    0.91 &    13.18 &   32.25 &    1.52 & --0.04\\
1441+252  &    --- &   --- &   --- &    21.74 &   24.00 &    1.10 &    18.83 &   31.85 &    1.02 &    --- &   --- &   --- & --0.31\\
1510--089  &    41.00 &   30.81 &    0.77 &    42.11 &   38.39 &    1.02 &    34.71 &   32.03 &    1.10 &    28.26 &   37.57 &    0.98 & +0.08\\
1553+113  &    --- &   --- &   --- &    23.67 &   18.45 &    0.65 &    21.48 &   19.31 &    0.59 &    --- &   --- &   --- & +0.02\\
1606+105  &    29.64 &   19.71 &    1.18 &    29.02 &   22.11 &    1.04 &    25.57 &   28.42 &    0.78 &    19.91 &   37.24 &    0.87 & --0.21\\
1611+343  &    17.59 &   10.56 &    0.69 &    17.27 &   13.63 &    0.69 &    27.45 &   20.04 &    0.58 &    15.10 &   29.54 &    0.62 & --0.30\\
1633+382  &    11.06 &   12.10 &    0.92 &    15.38 &   16.99 &    0.90 &    20.53 &   28.55 &    0.70 &     7.48 &   28.48 &    3.44 & +0.05\\
1641+399  &    10.29 &   10.36 &    0.95 &    12.74 &   11.72 &    0.79 &    13.90 &   19.20 &    1.02 &    24.22 &   39.65 &    1.42 & +0.02\\
1652+398  &     6.92 &    6.16 &    1.19 &     6.33 &    6.55 &    0.89 &     6.56 &    9.59 &    0.88 &    --- &   --- &   --- & --0.21\\
1730--130  &     9.70 &    8.83 &    1.30 &    15.95 &   12.54 &    1.05 &    17.50 &   23.27 &    1.31 &    17.01 &   28.96 &    1.04 & --0.25\\
1741--038  &    32.94 &   10.97 &    0.49 &    25.10 &   20.42 &    0.75 &    29.57 &   32.14 &    1.02 &    14.21 &   23.73 &    1.12 & --0.10\\
1807+698  &    11.41 &    8.48 &    1.51 &    17.47 &   11.52 &    0.74 &    22.70 &   23.53 &    0.94 &     8.53 &   18.94 &    1.13 & --0.06\\
1828+487  &     6.77 &    7.96 &    0.67 &     5.73 &    7.25 &    1.39 &    13.35 &   17.74 &    1.10 &    --- &   --- &   --- & --0.48\\
1830--211  &    --- &   --- &   --- &    25.07 &   27.19 &   --- &    28.57 &   23.16 &   --- &    --- &   --- &   --- & --0.47\\
1959+650  &    --- &   --- &   --- &    16.99 &   16.44 &    0.59 &    14.82 &   25.75 &    0.89 &    --- &   --- &   --- & --0.06\\
2200+420  &    34.39 &   24.78 &    0.64 &    41.56 &   27.52 &    0.55 &    42.50 &   34.45 &    0.68 &    24.18 &   24.52 &    0.76 & +0.12\\
2223--052  &    20.76 &   11.33 &    0.57 &    28.73 &   11.08 &    0.43 &    54.63 &   18.14 &    0.78 &    12.88 &   25.98 &    1.68 & --0.23\\
2230+114  &     8.01 &    5.90 &    0.59 &    18.92 &   18.17 &    0.85 &    30.38 &   31.57 &    0.96 &    13.95 &   41.15 &    0.97 & --0.01\\
2251+158  &    28.92 &   19.08 &    0.65 &    39.62 &   19.99 &    0.58 &    39.55 &   30.80 &    0.87 &    49.24 &   39.73 &    0.69 & --0.13\\
2344+092  &     8.73 &    8.47 &    1.35 &    14.35 &   10.66 &    0.65 &    22.24 &   24.63 &    0.87 &     8.74 &   21.10 &    0.77 & --0.12\\
2344+514  &    --- &   --- &   --- &     0.00 &   27.45 &    2.03 &     4.30 &   19.94 &    1.28 &    --- &   --- &   --- & --0.23\\
\hline
\end{tabular} 
\tablefoot{ The Table reports, for each source (Col. 1): the estimated $\overline{m}$ values (Cols. 2, 5, 8, and 11); the $SF_{1.5}^\prime$ values (Cols. 3, 6, 9, and 12); the estimated values of $SF_{1.5}^\prime/SF_{3.0}^\prime$ (Cols. 4, 7, 10, 13); the average spectral index between 8 and 24 GHz (Col. 14).}
\vspace{5cm}
\end{table}

\FloatBarrier
\newpage
\section{Light curves}
\label{sec:app2}

The light curves of the 47 sources monitored  within the ROBIN program.

\begin{figure}[h!]
\begin{center}
\centering
\includegraphics[width=1.\linewidth, height=1.\linewidth]{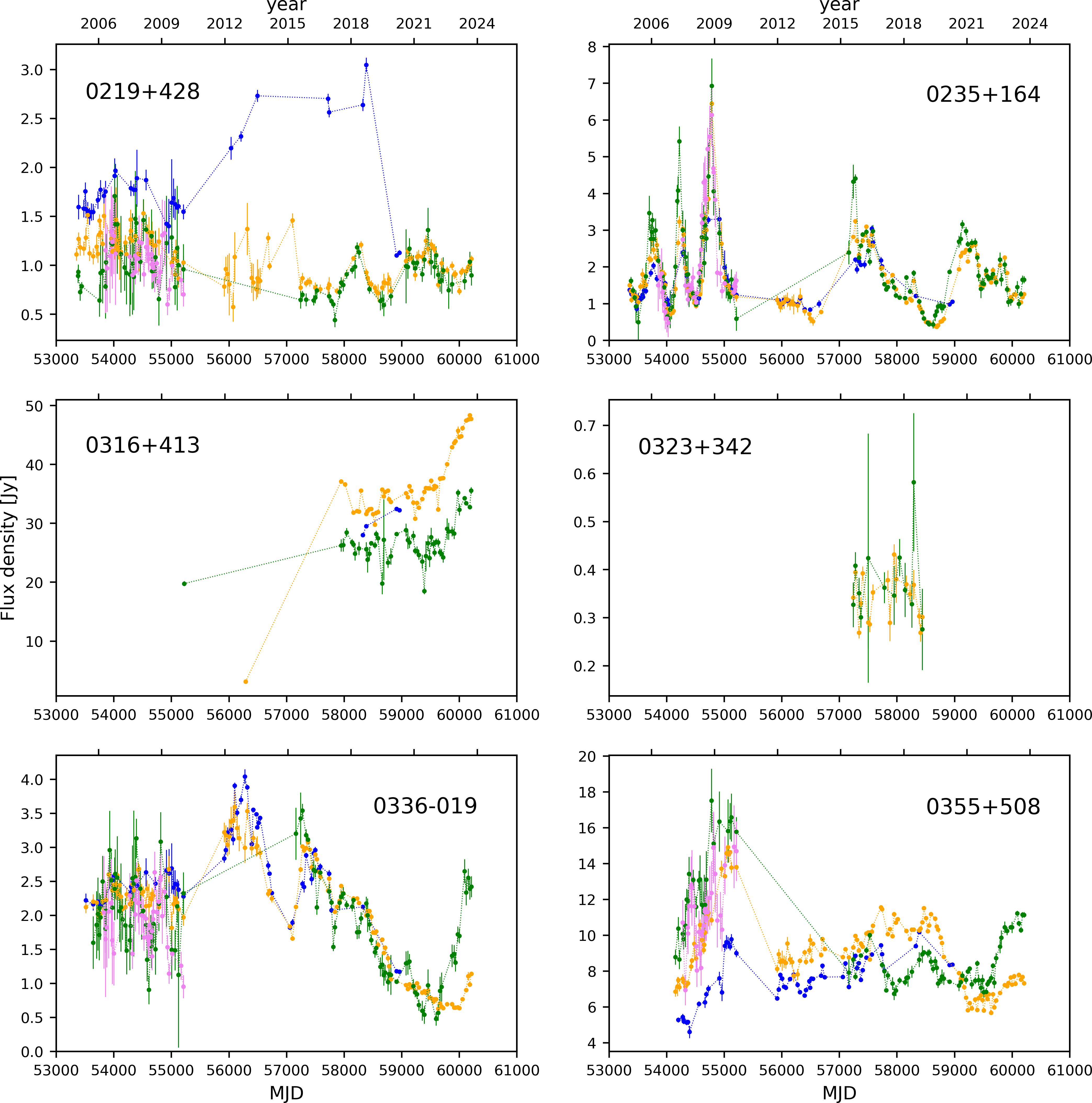}
\caption{In all plots, blue dots represent 5 GHz data, while orange, green, and magenta dots represent respectively 8, 24, and 43 GHz data.}
\end{center}
\end{figure}

\newpage

\begin{figure}
	\ContinuedFloat
\begin{center}
\centering
\includegraphics[width=1.\linewidth, height=1.\linewidth]{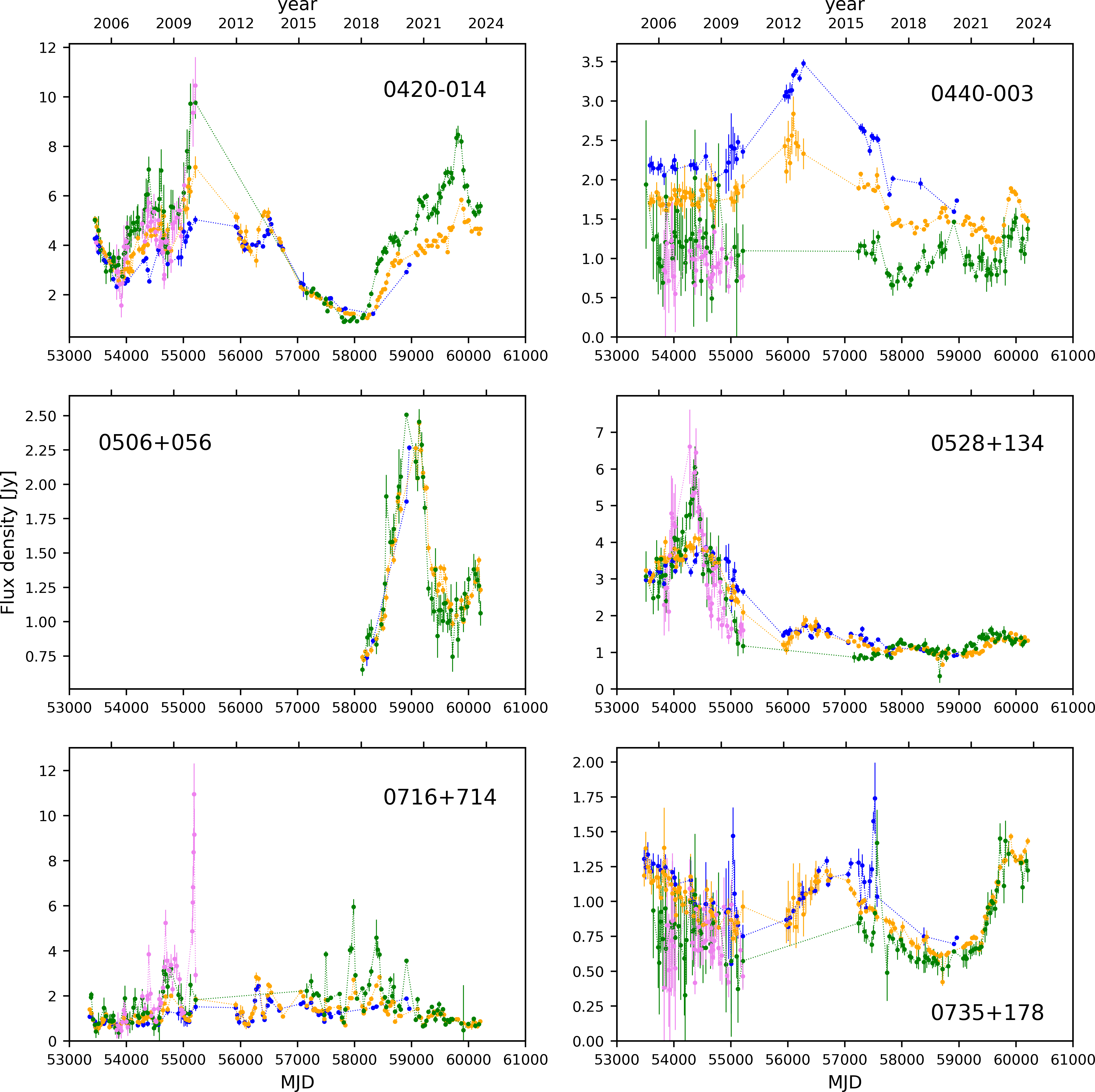}
\caption{continued.}
\end{center}
\end{figure}

\newpage

\begin{figure}
	\ContinuedFloat
\begin{center}
\centering
\includegraphics[width=1.\linewidth, height=1.\linewidth]{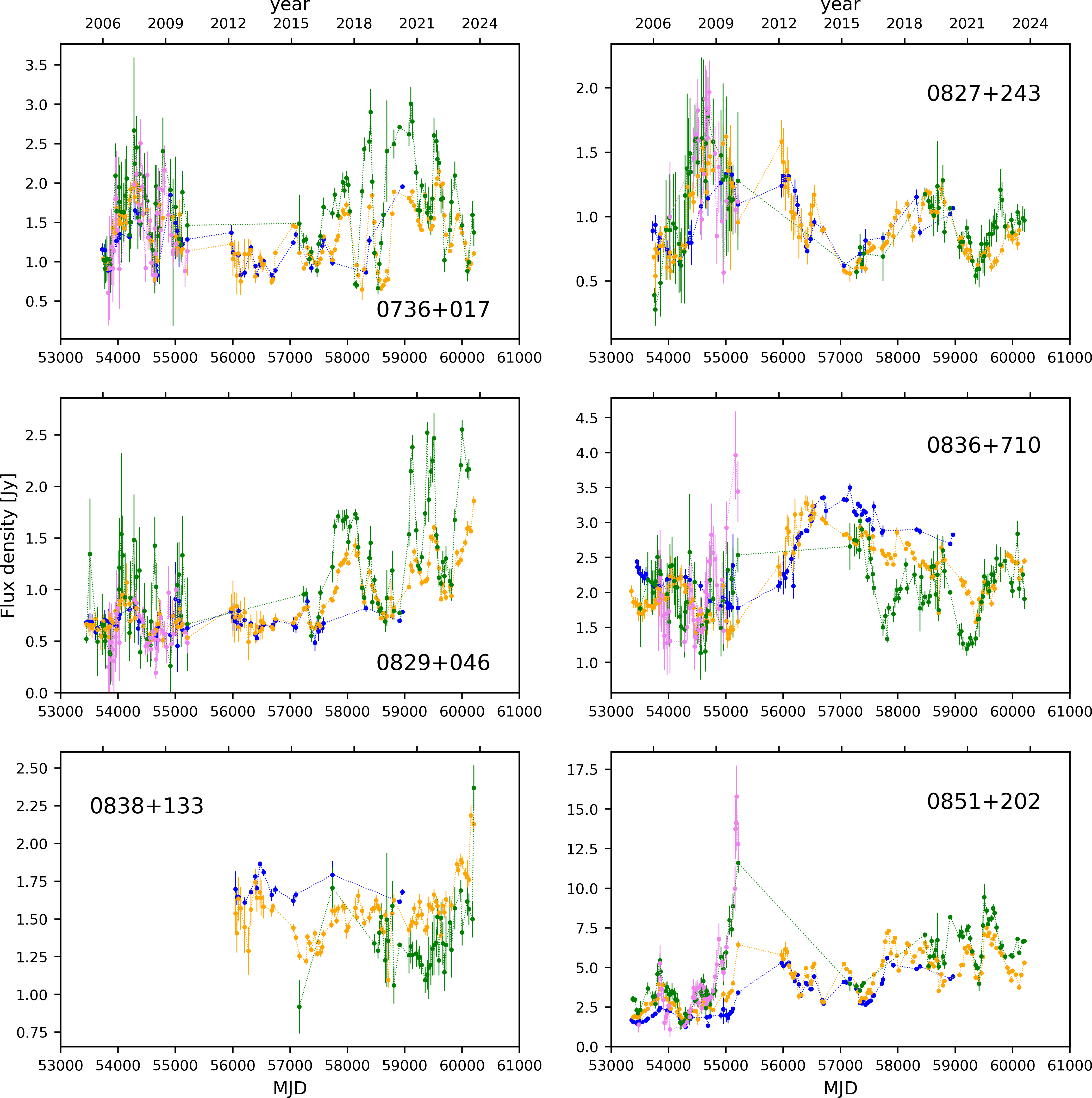}
\caption{continued.}
\end{center}
\end{figure}

\newpage

\begin{figure}
	\ContinuedFloat
\begin{center}
\centering
\includegraphics[width=1.\linewidth, height=1.\linewidth]{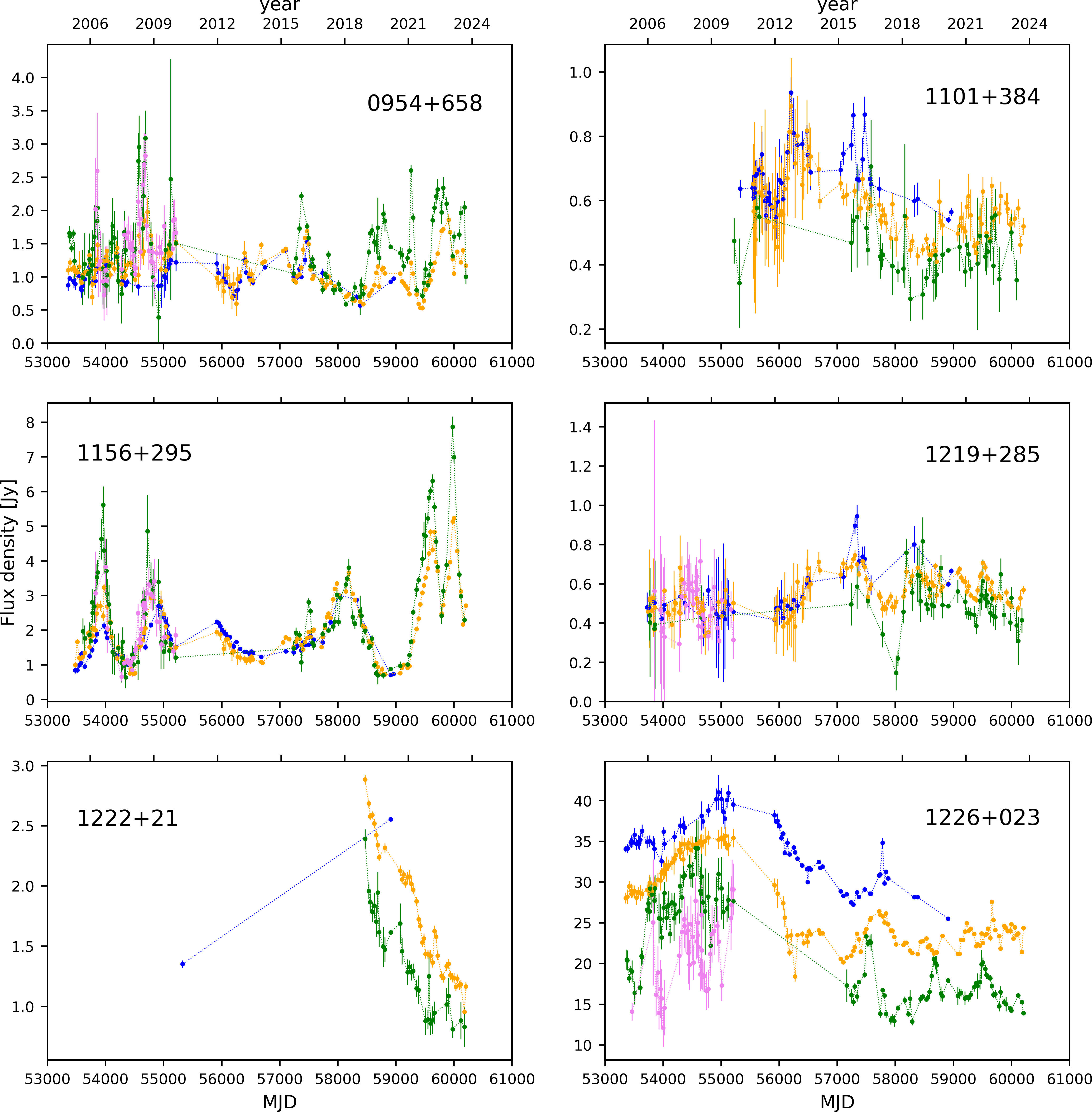}
\caption{continued.}
\end{center}
\end{figure}

\newpage

\begin{figure}
	\ContinuedFloat
\begin{center}
\centering
\includegraphics[width=1.\linewidth, height=1.\linewidth]{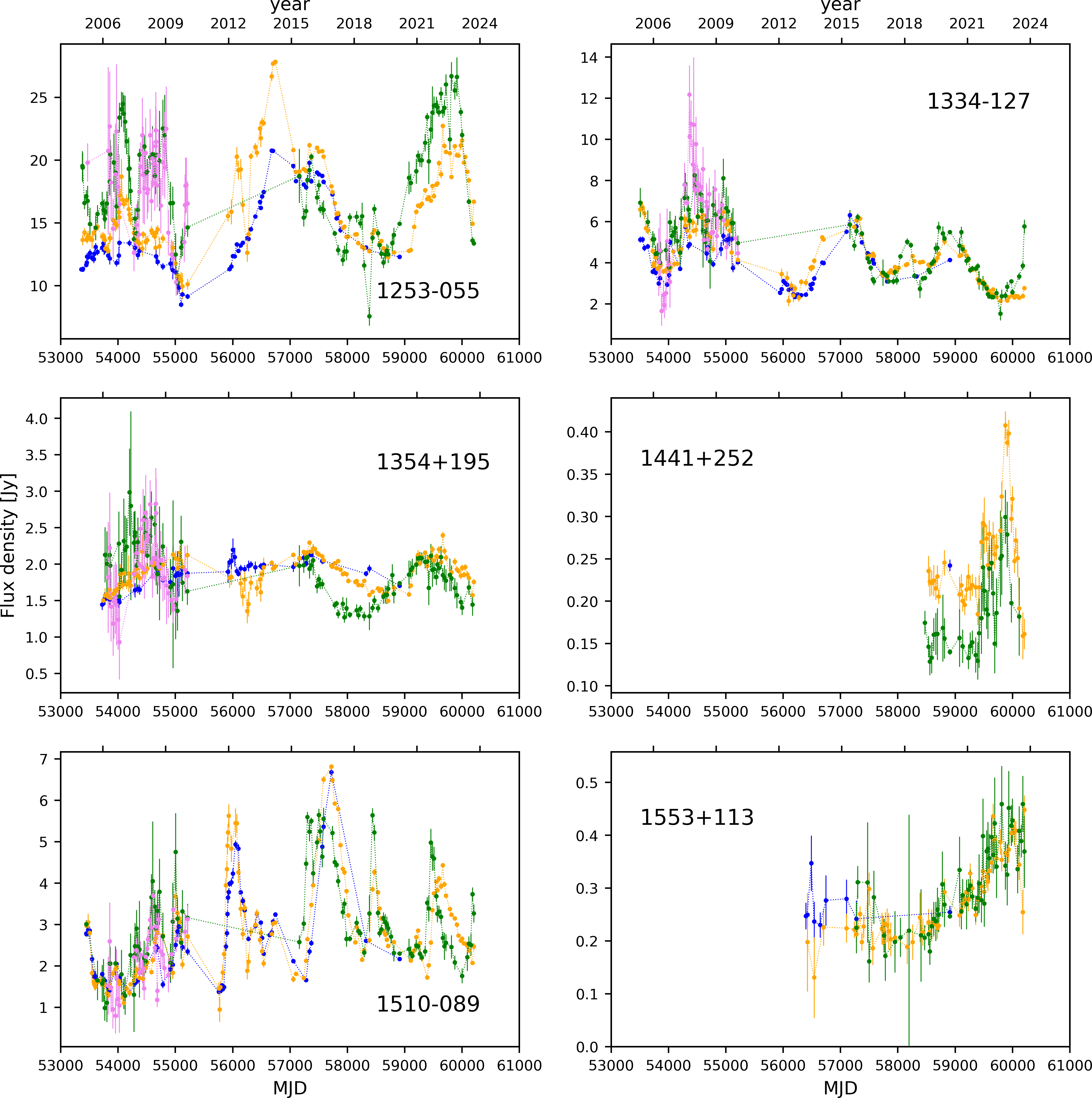}
\caption{continued.}
\end{center}
\end{figure}

\newpage

\begin{figure}
	\ContinuedFloat
\begin{center}
\centering
\includegraphics[width=1.\linewidth, height=1.\linewidth]{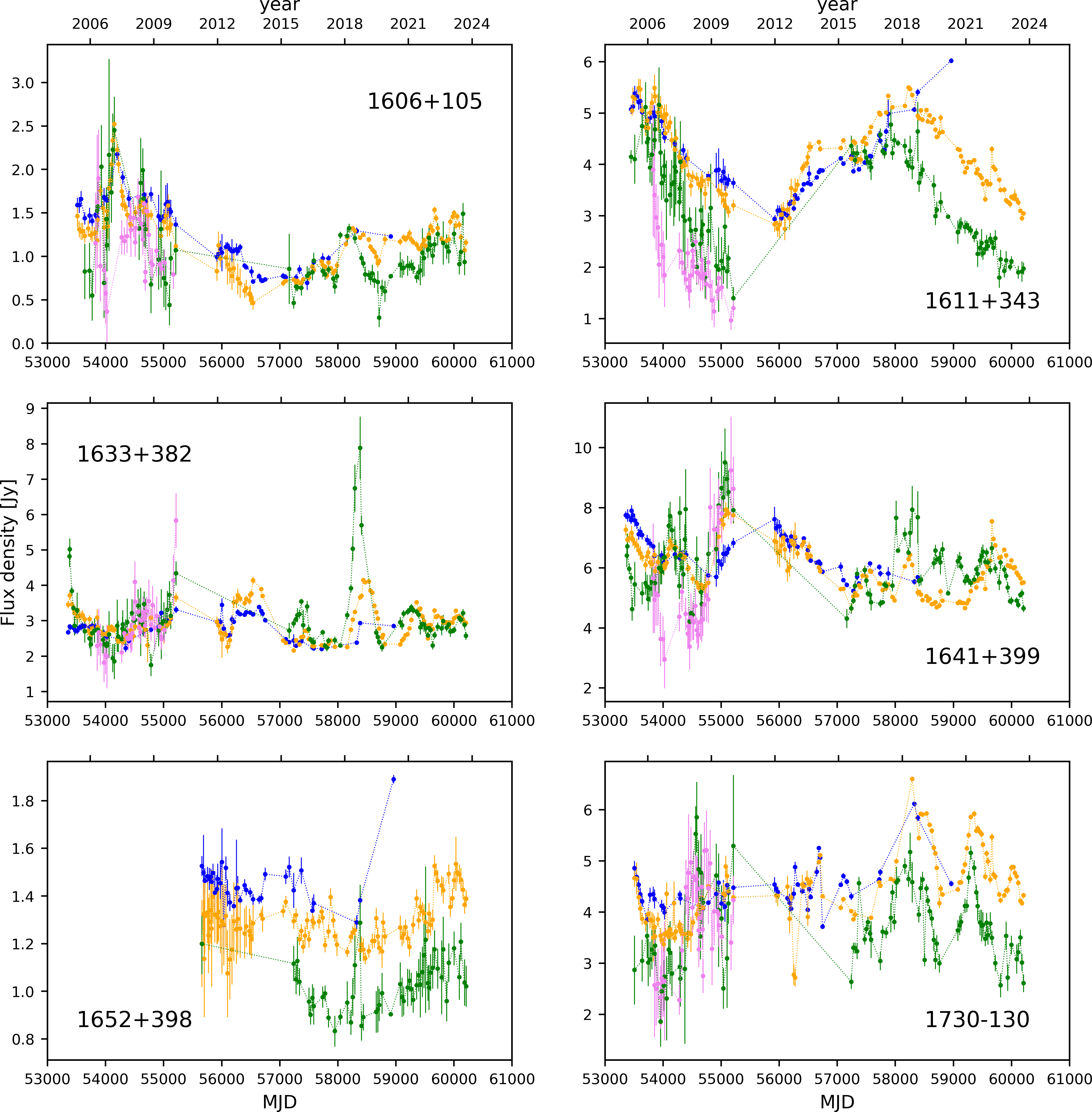}
\caption{continued.}
\end{center}
\end{figure}

\newpage

\begin{figure}
	\ContinuedFloat
\begin{center}
\centering
\includegraphics[width=1.\linewidth, height=1.\linewidth]{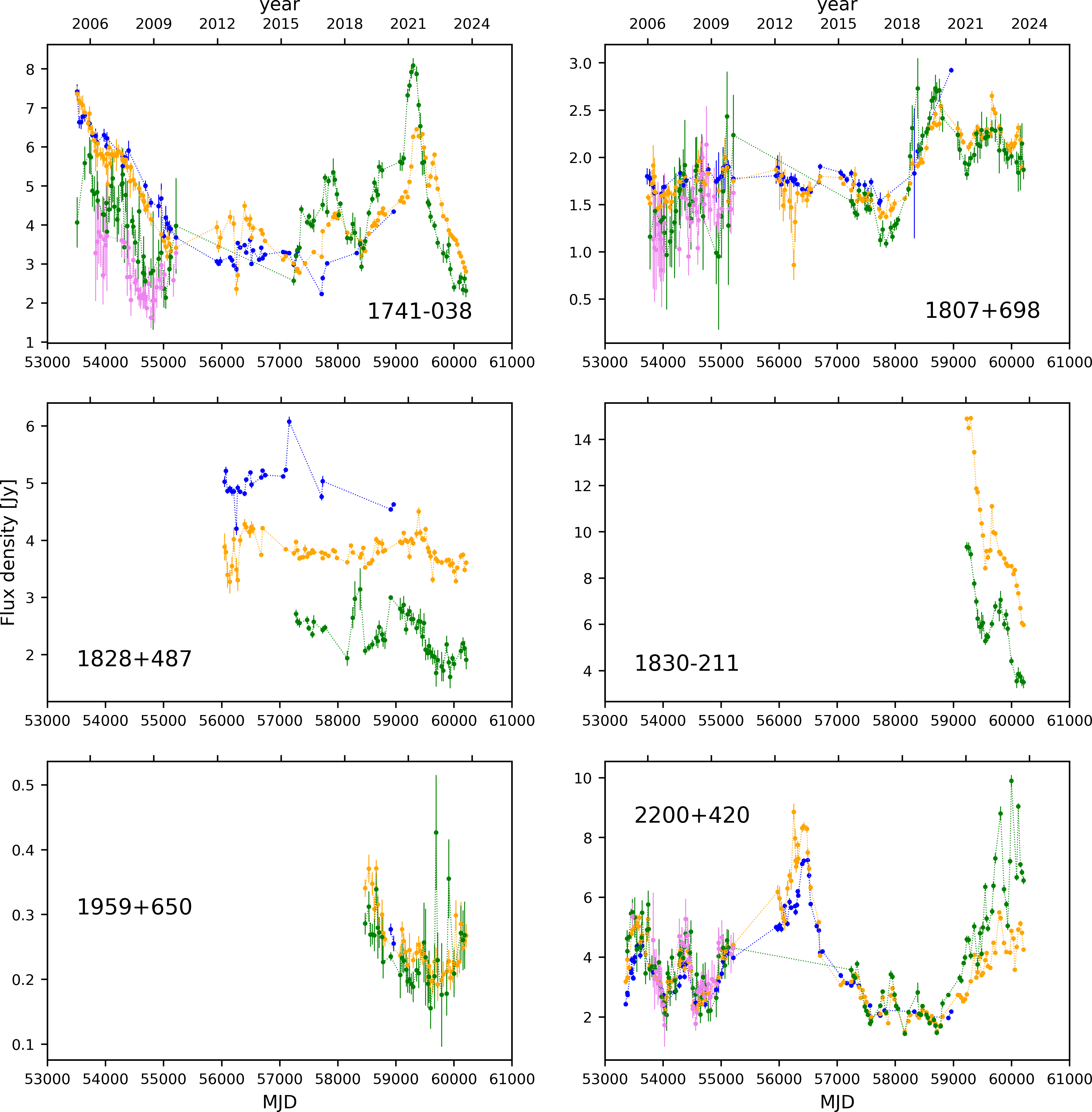}
\caption{continued.}
\end{center}
\end{figure}

\newpage

\begin{figure}
	\ContinuedFloat
\begin{center}
\centering
\includegraphics[width=1.\linewidth, height=1.\linewidth]{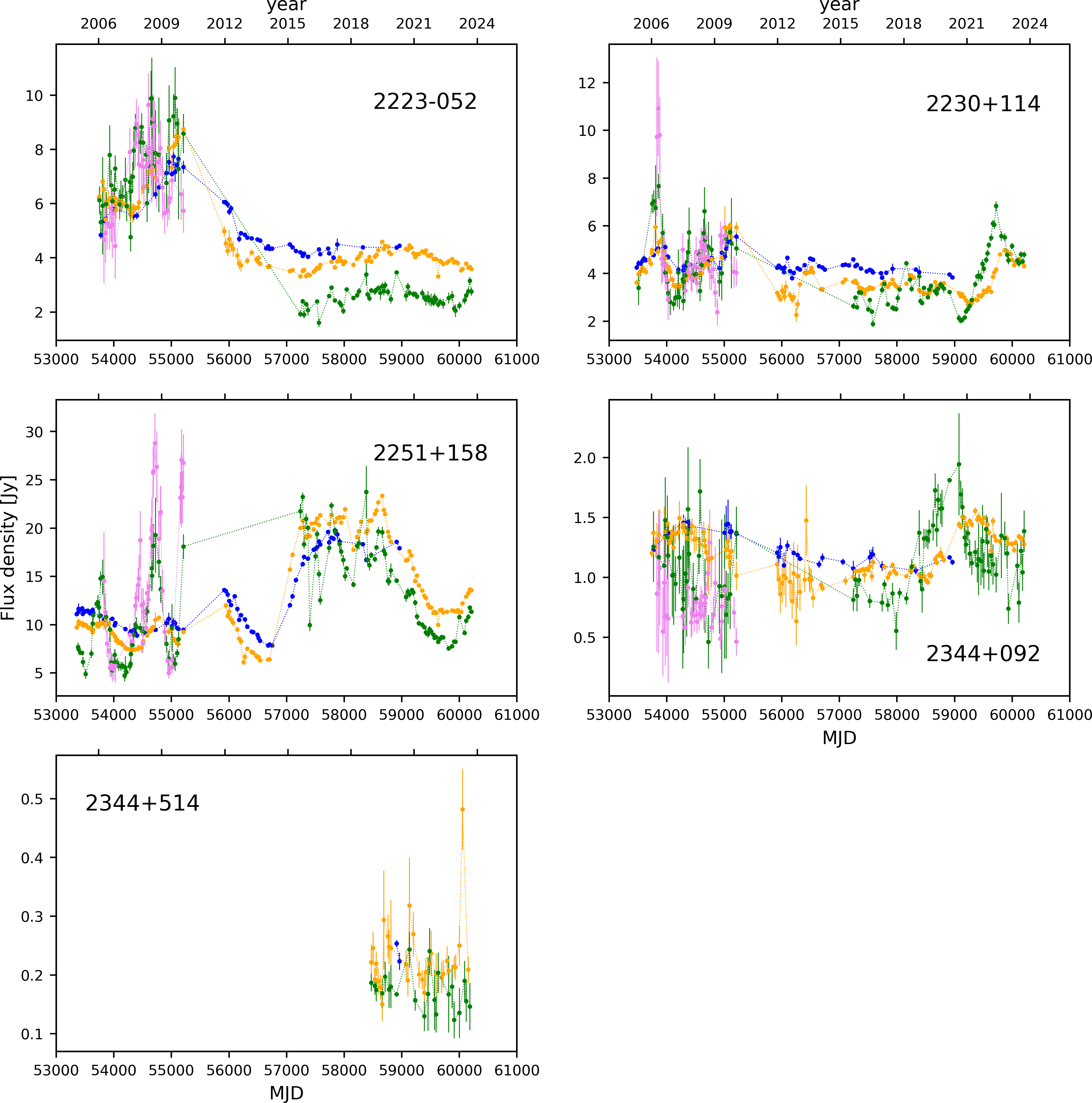}
\caption{continued.}
\end{center}
\end{figure}

\end{appendix}

\end{document}